# Synthesis of Dispersed Metal Particles for Applications in Photovoltaics, Catalysis, and Electronics

**Igor Sevonkaev**, **Vladimir Privman**, **Dan Goia**

Center for Advanced Materials Processing, Department of Chemistry and Biomolecular Science, and Department of Physics, Clarkson University, Potsdam, NY 13699, USA

### Abstract

In colloid and nanoparticle chemistry, particle size, shape, crystallinity, surface morphology and composition are controlled by employing the mechanisms of burst nucleation, diffusional growth, aggregation, or their combinations. Here we review and survey practical examples of recently developed methods for preparing metal colloids and nanoparticles for industrial applications such as photovoltaics, catalysis, and consumer electronics. We discuss relevant theoretical models, many of which are general, and identify growth mechanisms that play a major role in other systems and applications as well.

**Introduction**

Metallic particles with carefully tailored properties are widely used in many areas of technology and medicine. Metallurgy, catalysis, consumer electronics, and pigments are already mature and well established fields requiring large quantities of metal powders with strict specifications. In contrast, the success of many applications in bio-medical, optoelectronic, and other emerging fields of high technology depends on the availability of metallic particles (MPs) with unique and tunable novel properties. The ability to control size, internal structure/morphology, composition, shape, surface characteristics, as well as the distribution of these and other properties of MPs is crucial for modern technology. To do so, it is essential that the mechanisms of particle formation are well understood and the key process parameters tightly controlled.

Usually the most important parameter considered when selecting a product for a given application is particle size. For example, single digit nanometer (2–6 nm) metallic particles of noble metals (Pt, Pd, Rh, etc.) are a necessity in catalytic applications as they offer very high specific surface areas and thus a large fraction of catalytically active surface atoms [1-5]. While such particles may be used in some cases in the form of dispersions, in most catalytic applications they are supported on larger substrates. The two cases are illustrated by Fig. 1A and 1B (note that all the figures are located at the end of this preprint: pages 34–50), which show pure platinum nanoparticles (~ 3 nm) prepared by refluxing a solution of $(EA)_2[Pt(OH)_6]$ in ethanol and the same nanoparticles deposited on a carbon substrate. A similar rationale is at work in the case of antimicrobial applications [6-8] of very small silver nanoparticles. Figure 1C, for example, shows highly dispersed Ag nanoparticles with an average size of ~ 10 nm obtained [7] by heating a solution of silver salicylate in diethylene glycol in the presence of Daxad 11G.

For applications relying on the optoelectronic properties of particles, a small size (diameters less than ~ 100 nm) is a necessary but not sufficient condition to trigger the electron oscillations responsible for the manifestation of localized surface plasmon resonance absorption bands [9]. Indeed, excellent particles dispersibility and uniformity, are also ~~among the~~ necessary attributes to have well defined plasmon bands. Gold nanoparticles obtained by reducing tetrachloroauric acid with aminodextran, see Fig. 1D, represent a good example of a suitable plasmonic material [10, 11]. The presence of plasmon bands is the basis for bio-medical [12] and both silicon [9] and polymer thin film solar cell [13-15] applications. Specifically in the latter, plasmon bands can dramatically enhance the absorption of light. Such highly dispersed uniform metallic nanoparticles are usually precipitated from homogeneous solutions via a mechanism consisting of a short nucleation burst followed by a limited diffusional growth [16]. These and other growth mechanisms will be addressed in later sections.

Metallic particles ranging from 100 nm to few microns are typically used in the manufacturing of most conventional electronic devices [6, 17, 18] and crystalline silicon solar cells [9, 19-21]. Primary applications for MPs of this size range, have been in sintered metallic structures (conducting tracks) of 10-100 μm. Utilization of smaller particles in such instances would unnecessarily complicate the device manufacturing due to their excessive reactivity during the sintering process. Preparing uniform highly dispersed MPs in this size range is a challenging task as well. In principle, particles could be prepared either by growing via addition of metal atoms or by controlled aggregation of nanocrystals. The former approach requires slow reduction rates (to prevent other nucleation bursts) and usually leads [22-25] to the formation of highly crystalline particles, see Fig. 2. The latter, aggregation mechanism, is a process which occurs rapidly and yields [26-28] larger polycrystalline particles, see Fig. 3.

For a comparable size, the internal structure of the particles may strongly affect sintering and conductive properties of the resulting metallic structures. Since the densification of highly crystalline particles predominantly involves mass transfer between them, sintering would typically require high temperatures and longer dwell times. Therefore, materials of choice for forming dense polycrystalline metallic structures are those in which individual particles retain their crystallinity, i.e., those used in plasma display panels, co-fired metal/ceramic assemblies, and fired conductors/inductors [29-33]. In the case of the initially polycrystalline particles, however, a significant intra-particle restructuring occurs simultaneously with the inter-particle mass transport responsible for densification. Consequently, they typically sinter more rapidly and densify at lower temperatures, which makes them well-suited for silicon solar cells applications.

It is noteworthy that both mechanisms are also at work in the deposition of metallic coatings on surfaces of particulate substrates. Encapsulation of core particles of different size, shape, and composition in metallic shells has major implications in both catalysis and electronics, because core-shell structures provide significant performance enhancement at lower cost [1, 4, 34]. For example, continuous crystalline shells consisting of several atomic layers of platinum, see Fig. 4A, have been deposited by orderly diffusional growth/deposition of shell atoms on crystalline gold cores to obtain high efficiency electrocatalysts for proton exchange membrane fuel cells (PEMFC). The resulting core-shell structure displays a mass catalytic specific activity of Pt, several times higher than that of conventional materials based on Pt nanoparticles [35]. The deposition of thicker metallic shells ( > 50 nm) onto less expensive or chemically reactive particulate substrates is a common practice in the electronics industry. Formation of these layers by aggregation of smaller entities is a more convenient approach than the diffusion controlled growth which is slow and expensive. Figure 4B depicts polymer spheres coated with uniform polycrystalline shells of Ni [36]. Once coated with a thin (10-20 nm) layer of gold, such beads are used as conductive filler in anisotropic conductive adhesives (ACA), which are essential materials in the manufacturing of liquid crystal displays (LCD).

Internal composition of MPs also depends on the interplay of particle formation mechanisms and has a significant impact on their performance in catalytic and electronic applications. For instance, oxidation and sintering of AgPd particles play major roles in the manufacturing of multilayer ceramic capacitors (MLCC). Both phenomena are affected not only by the atomic ratio of the two metals but also by their distribution inside each particle. Indeed, alloyed AgPd particles are more resistant to oxidation than particles consisting of a Ag core and a Pd shell, for example. By carefully controlling the experimental conditions, bi-metallic particles with similar size and shape but different internal structure can be precipitated from homogeneous solutions. When the $Ag^+$ and $Pd^{2+}$ species are co-reduced in strongly acidic medium (large excess of nitric acid) with ascorbic acid, the two elements are deposited at similar rates forming spherical aggregates consisting of small (10-14 nm) AgPd alloy crystallites, see Fig. 5B. In contrast, the reduction the ammonia complexes of the two metals with hydrazine hydrate at elevated pH (11.0 ± 1.0) favors the precipitation of Ag first (the less stable ammonia complex) followed by the deposition of Pd as an external shell, see Fig. 5A. As the alloy particles tend to oxidize at a lesser extent, they are better suited for building capacitors with a higher number of layers and thus higher volume capacitance [37, 38].

The shape of MPs is another factor affecting performance in catalytic, optical, and electronic applications. It has been shown, for example, that anisometric Pt nanoparticles with extensive (1 1 1) facets tend to have higher catalytic activity [39]. Anisometry also has a large impact on the position of the plasmon band for nanorods [40, 41] and nanoplatelets [42]. Finally, in the electronic industry flakes/platelets of conductive metals are extensively used for obtaining conductive structures in membrane touch switches, conductive adhesives, and electromagnetic interference shielding applications. In all these cases the conductivity of the final structures relies on the tunneling of electrons between particles. While anisometric metallic particles are obtained mostly by mechanical deformation of isometric particles, it is possible to control the formation mechanisms of MPs during precipitation to yield platelets of improved uniformity and shape. For example, uniform crystalline Ag and Ag/Pd nanoplatelets, see Fig. 6, were obtained by controlling the nucleation and growth processes during the reduction at room temperature of the respective metal nitrates with ascorbic acid in concentrated nitric acid solutions containing Arabic gum [25, 43]. Two elements of this particular system favor the anisotropic growth of silver. The first is the slow release of electrons from the ascorbic acid molecule, which favors diffusional deposition of Ag atoms. The second is the strongly oxidizing environment provided by the excess nitric acid, which causes more pronounced re-dissolution of silver deposited on specific crystal facets. The balance of the two processes (one additive, one subtractive) leads to the formation of anisometric particles.

In some applications, control of the overall particle properties and their uniformity is not as crucial as control of surface properties. High quality crystalline-face substrates may be desirable, for example, for shell-core electrocatalysts [22, 24]. For instance, Fig. 7 illustrates such "highly crystalline" (in the sense detailed later) Ni particles, grown by diffusional transport of solutes [24]. The synthesis consists in a slow reduction of nickel basic carbonate in diethylene glycol at elevated temperature (210-220 °C). The nucleation is controlled by the "seeding" method, a step in which a small amount of a noble metal (Ir, Pd, Pt) salt is reduced first at lower temperature (100-120 °C). The nickel is subsequently deposited on these seeds by adjusting the temperature of the reaction in a range where the reduction of $Ni^{2+}$ species is very slow. These particles are not necessarily highly uniform in their sizes or shapes, but offer improved surface-face morphologies needed for catalysis and electronics [22]. Specifically, the particles have practically no defects at their surfaces, providing improved conditions for further epitaxial deposition of shell materials.

These examples, along with many others, demonstrate the importance of understanding the processes involved in controlling the formation of nanosize particles and their manipulation in order to design and manufacture novel materials with enhanced functionality. Most experimental findings reported to date have been generated by a "trial and error" approach. As a result, only partial theoretical understanding and predictive capabilities have been gained regarding the role of processes such as diffusion, aggregation, nucleation, dissolution, restructuring, and interplay of these and other kinetic mechanisms in determining the properties of the synthesized products. In the following sections we address several theoretical approaches, some, but not all of which have been checked against the available experimental data. The latter are typically limited to the observation of the final products at the largest length scales. Indeed, quantitative experimental information on the growth kinetics especially for short times, at the few-atom and nanosize scales is rarely available.

In the following sections, we review our present understanding of the modeling approaches to various processes and steps of nanoparticle and colloid synthesis. While we consider the dispersed-particle synthesis typical for colloid chemistry, we also reference recent attempts to apply these techniques for on-surface growth, which promise future applications in the context of topics covered in the present Special Issue. Our review also highlights the widely varying degree to which availability of data has allowed comparison of the general models of growth processes to the relevant experiments.

**Approaches to Modeling the Mechanisms of Control of Particle Synthesis**

Several multi-scale kinetic processes are involved in particle synthesis. These processes could involve transport of matter, nucleation, growth, aggregation, surface restructuring, and detachment [44-48]. The "matter" can be transported on all scales: from atoms, molecules or ions (which could in turn be undergoing chemical reactions with each other and with solution species), to nanosize and larger objects, including already formed clusters/particles/structures. Approaches to modeling are therefore numerically intractable unless specific kinetic mechanisms can be singled out that capture the size, shape or other specific property the control of which is of interest. In the following sections we review models [44, 49-65] of processes of burst-nucleation of nanocrystals [46], and of diffusional growth of nanoparticles [22, 24], then include the secondary process of nanocrystals' diffusional aggregation into polycrystalline colloids [10, 11, 27, 28]. The models provide semi-quantitative description of particle size selection. Understanding of particle shape and morphology emergence [40, 41, 43] is less developed theoretically, with only the first modeling results recently published [44, 49-51, 63], which will not be reviewed here.

Narrow particle size distribution has been a traditional goal of colloid-chemistry synthesis approaches [62, 66, 67]. Recent challenges have been related to the demand for small particle sizes. Size (and shape, etc.) control at the nanoscale differs from that utilized for micron and submicron particles, whereas experimental data on the time dependence of the nanoscale growth stages is usually not obtainable.

The spatial transport of solute/suspension matter is typically diffusional [24, 64]. The constituent units from which particles are formed, frequently called monomers or singlets, in nanosize synthesis are solute species: atoms, ions, molecules. However, for the process of formation of polycrystalline colloids, the singlets can be precursor primary-particle nanocrystals, themselves formed by burst nucleation [64, 68]. Synthesis processes involve matter dissolved/suspended in aqueous or non-aqueous medium. They can be externally controlled not only by the initial supply of reactants, but by varying the chemical and physical conditions during the process. Furthermore, matter can also be introduced externally during the growth, ranging from chemical-reaction release of atoms, to seeding.

Size distribution can be defined in terms of the number of constituent units, $s$. It is desirable to have a narrow peak centered at $s_{\text{peak}}$, as the particles grow to final sizes: Fig. 8. Most standard transport processes during synthesis are in the regime such that the average particle size at the peak, $s_{\text{peak}}(t)$, grows with time, $t$. However, they also typically cause broadening of the peak. These properties are rather general and apply to several processes which are driven by diffusional transport of matter, including cluster-cluster aggregation and cluster ripening due to exchange of monomers. Indeed, larger particles have bigger surface area for capturing additional matter, and their surfaces have less curvature and a smaller fraction of edges, corners, etc. On average, this results in stronger binding/less detachment of monomers, etc. The larger-$s$ side of the peak (see Fig. 8) advances faster than the smaller-$s$ side. Thus, the peak generally broadens during growth [24, 65].

Therefore, "natural" growth processes in most situations yield particles with broad size distributions. Similarly, growth is also accompanied by surface fluctuations which result in random/nonuniform/broadly distributed shapes and other properties. Special approaches and selective growth regimes must be identified and maintained to get some or most particle property distributions sharply peaked. "Narrowly distributed" is frequently called "monodispersed" in the colloid literature. To obtain a narrow size distribution, one could limit the growth of the large-$s$ particles by forming them inside micelles or inverse micelles [69, 70]. Seeding, i.e., growth on top of earlier prepared cores [22, 24, 71] (see Fig. 7) has also been extensively used.

Other approaches rely of the specific growth-kinetics choices. Among these we consider [59, 60, 64] burst nucleation of nanocrystals rapidly growing in a supersaturated solution. Narrow size distribution is then accomplished by that the small-$s$ side of the peak (see Fig. 8) is eroded by the thermalization of sub-critical clusters. However, other processes broaden the distribution after the initial nucleation burst, limiting this mechanism to the nanosize growth stage.

A two-stage colloid growth mechanism [58-60, 62] for polycrystalline colloids, can yield narrow size distributions. Precursor nanocrystals of concentration $C(t)$ are burst-nucleated and serve as monomers for secondary aggregation to form colloids (see Fig. 8). The peak then grows to large sizes fast enough, that it is not significantly broadened. This only occurs provided that the primary and secondary processes yield the time-dependence of $C(t)$ such that no significant "shoulder" develops for small $s$ (see Fig. 8). We will discuss this approach in a later section. For nanoparticle growth, there have also been approaches [72, 73] based on stepwise addition of batches of atomic-size monomers, similar to the aforementioned control of $C(t)$.

Let $N_s(t)$ denote the density distribution: number of particles containing $s$ monomers, per unit volume of the suspension, at time $t$. Except for very small values of $s$, this distribution is treated as a function of continuous $s$ (see Fig. 8). The monomer ("building block") concentration, however, is frequently separately controlled,

$$C(t) = N_1(t) ,$$
$$dC(t)/dt = \rho(t) - \ldots .$$
(1)

They can be introduced practically instantly or gradually, or synthesized by another process, at the rate $\rho(t)$ per unit volume. At the same time they are consumed by aggregation into small clusters in the "shoulder," as well as by being captured by large clusters (particles) including those in the main peak. These latter processes yield various negative terms in the rate expression, indicated by … in Eq. (1).

The initial emergence of the main peak also requires explanation. It is formed naturally in burst nucleation of nanocrystals due to the difference in the kinetics of small and large clusters. For larger particles (colloids), the peak formation is a byproduct of cluster-cluster aggregation at the early growth stages. Seeding is of course a useful mechanism to initiate the peaked size distribution.

Control of particle shape distribution for uniformity, is much less understood theoretically [44, 49-51, 63] for the relevant type of growth: fast, nonequilibrium. Shape-selection mechanisms are not unique and depend on the specific situations. Recent studies [44, 49-51, 63] have suggested that growth without development of large internal defect structures, can yield well defined shapes with crystal faces similar to those in the equilibrium crystal form, but with different particle proportions. These shapes and surface morphologies persist for a range for particle sizes during their growth. Such ideas, not reviewed here, have also recently been applied [50, 51] to study the morphology of nanostructured surfaces.

**Burst Nucleation**

Burst nucleation is a process [59, 60, 64, 74, 75] induced by a significant supersaturation of singlets: atom, molecules or ions, i.e., the solute-species monomers for growth. For nanosize particles (growing clusters) containing $n$ monomers, a critical cluster size, $n_c$, is identified within a Gibbsian approach to nucleation: Fig. 9. It is assumed that the "supercritical," $n > n_c$ clusters grow irreversibly. They capture diffusing solute monomers. The size distribution of the "subcritical" clusters, those smaller than $n_c$, which are also called embryos, in the "shoulder" in Fig. 9, is assumed instantaneously thermalized.

Burst nucleation sets in at time, $t = 0$, when monomers are introduced or produced via a chemical reaction at the initial concentration, $c(0)$, significantly exceeding the equilibrium concentration, $c_0$. Thermal fluctuations cause formation of embryos, but the free energy of the forming particles has a barrier peaked at $n_c$, due the cost of the cluster surface formation. Indeed, the surface matter is less bound than that in the interior of the cluster. While the internal restructuring dynamics of few-atom clusters and their equilibration with the surrounding solution are not well understood, these processes are assumed to be practically instantaneous for clusters smaller than $n_c$. Therefore, the $n < n_c$ embryos are approximately thermally size-distributed according to the Gibbs-type free-energy

$$G(n,t) = -nkT \ln[c(t)/c_0] + 4\pi a^2 n^{2/3} \sigma ,  \qquad (2)$$

where $k$ is the Boltzmann constant, $T$ is the temperature, $\sigma$ is the effective surface tension. Here $a$ is defined such that the radius of an $n$-atom embryo is $an^{1/3}$ and can be estimated by requiring that $4\pi a^3/3$ equals the unit-cell volume per monomer in the bulk material. The free-energy has its maximum at $n_c$, see Fig. 10,

$$n_c(t) = \left(\frac{8\pi a^2 \sigma}{3kT \ln[c(t)/c_0]}\right)^3 . \qquad (3)$$

The first, "cluster interior" or "bulk" term in Eq. (2), proportional to the volume (to $n$; we ignore small-$n$ corrections), is negative (because $c > c_0$), favoring cluster growth. Its distinctive feature for burst-nucleation is the logarithmic dependence on the monomer concentration, $c(t)$, which introduces the explicit time dependence of $G(n,t)$ and $n_c(t)$. This term accounts for the loss of "entropy of mixing" of noninteracting solutes as they are bound in the cluster interior. The denominator, $c_0$, in the logarithm, provides a reference to the free-energy gained by the binding of solutes. The second, "surface free-energy cost" term, proportional to the surface area, $\sim n^{2/3}$, is positive, i.e., it disfavors growth of clusters.

Recall that clusters are assumed instantaneously thermally distributed for $n < n_c(t)$. For $n > n_c(t)$, clusters grow irreversibly, by capture of solutes. All the above assumptions, including the surface-bulk free-energy structure, Eq. (2), are standard for homogeneous nucleation. The distinctive property of the process of burst-nucleation is that the bulk term is explicitly dependent on $c(t)$ and therefore varies with time. As mentioned, the critical cluster size and the height of the nucleation barrier are therefore time-dependent.

Obviously, nucleation theories of the type considered here involve numerous simplifications and assumptions, focusing on the size distribution but ignoring variations in particle shapes and other structural properties. An idealized, compact spherical shape is assumed for all the clusters. Surface details and other properties, such as, for crystals, symmetry faces, edges, corners, etc., effect matter transport and binding in the structure. Even within the spherical shape approximation, the "surface tension" parameter, $\sigma$, for instance, depends on the radius (surface curvature). Geometry and morphology variations/corrections are ignored is the nucleation theory because of the computational difficulties of treating multi-variable distributions. Furthermore, quantities such as $\sigma$ for nanosizes, are presently understood and experimentally quantifiable only to a very limited extent, e.g., [76]. For example, $\sigma$ in Eqs. (2)–(3) is usually not available from direct measurements. It has either been treated [53, 54, 57, 62] as an adjustable parameter or set to the known bulk-material surface tension, $\sigma_{\text{bulk}}$.

Another problem has been that experimental observations only yield size and other information for the final products, whereas time-dependence data are rarely obtained, e.g., [46, 53], especially for nanoparticle growth. Interestingly, these limitations apply [79] even for protein crystal growth, for which monomers are very large as compared to, for instance, metal atoms. Models have to be used to relate the dynamical behavior at the level of monomers to the observed properties. Recently, such an approach to model validation, by utilizing multi-scale numerical calculations for industrial processes has been advanced in [77, 78].

Let $P(n,t)$ denote the cluster size distribution at time $t$. This notation is similar to that of $N_s(t)$ introduced earlier in connection with Fig. 8 and reserved for later use. Here also, for small values of $n = 1, 2, \ldots$ the quantity $P(n,t)$ is discrete, and we single out $c(t) = P(1,t)$, However, for larger $n$, the distribution is usually replaced by a continuous function $P$ such that $P(n,t)dn$ gives the number density for particles of sizes between $n$ and $n + dn$. Burst nucleation is initiated by introducing or chemically generating monomeric matter at time 0, with $P(n,0)$ concentrated at very small $n$ values, and specifically, $c(0) \gg c_0$. The induction time involved in establishing the initial distribution is usually rather small (we just set it to 0), but the model can be extended to slow processes of creating the typically very large supersaturations used in syntheses situation considered here. For later times, $c(t)$ decreases from its large initial value towards $c_0$ and as a result the logarithmic (entropic) bulk term in Eq. (2) decreases in magnitude. Thus the barrier for nucleating new supercritical clusters grows with time, whereas the particle size distribution then evolves into the late-stage form [59, 60, 64] sketched in Fig. 9. Recall that the subcritical embryos are assumed thermalized on time scales faster than other kinetic processes,

$$P(n, t) = c(t) \exp[-G(n, t)/kT], \quad \text{for } n < n_c(t). \tag{4}$$

The approximate, but of course not the actual particle-size distribution in burst-nucleation is discontinuous at $n_c(t)$, see Fig. 9, because the growth of clusters "going over the barrier" at $n_c$ at the rate $\rho(t)$ per unit time, per unit volume, is assumed to proceed irreversibly. The growth rate for $n \geq n_c$ can be modeled by $K_n c P(n, t)$, where the kinetic coefficient within the simplest modeling approach can be selected [62] as

$$K_n = 4\pi (an^{1/3})D, \tag{5}$$

with $D$ the diffusion coefficient of monomers, and $an^{1/3}$ an estimate of the cluster radius (more elaborated expressions are possible and some will be addressed later). The prefactor $K_n c$ in the growth rate expression is the Smoluchowski rate for the irreversible capture of diffusing monomers of density $c$ by spherical clusters of that radius. Specifically, the nucleation rate per unit volume is

$$\rho(t) = K_{n_c} c P(n_c, t) = K_{n_c} c^2 \exp[-G(n_c)/kT]. \tag{6}$$

Note that $D$ in a dilute solution of viscosity $\eta$, is frequently estimated as $D \approx kT/6\pi\eta a$, up to geometrical correction factors relating the effective solute radius $a$ to the hydrodynamic radius.

If we replace $c(t)$ by $c(t) - c_0$ in the Smoluchowski rate constant, then growth will stop as $c(t)$ approaches $c_0$ at large times. One can show [55] that this approximately accounts for the detachment of matter, provided we ignore curvature and other surface-shape and structure effects. If they were accounted for, the latter effects would yield a variable (curvature-dependent) effective "equilibrium concentration" for different cluster sizes and, with monomer detachment, to the process of Ostwald ripening [85] by exchange of monomers between clusters. This and other possible coarsening processes, such as cluster-cluster aggregation [80, 81] are typically slower than burst nucleation [59, 60, 62, 64]. A kinetic equation, with $c(t) - c_0$ in the rate, for the irreversible burst-nucleation growth of clusters with $n > n_c(t)$ was introduced and analyzed in [9]. Note that burst-nucleation alone leads to a linear growth of $n_c(t)$, which also happens to be the $n$ value at which the distribution is peaked (see below) at large times [59, 60, 64], as shown in the inset in Fig. 9. However, the slope turns out to be very small [82] in typical experimental situations. Thus, particle growth would practically stop. However, for later times the other coarsening processes will take over (meaning that the burst-nucleation approach approximations break down). They typically not only grow, but also substantially broaden the size distribution as time goes by.

Internal restructuring plays an important role in the structural evolution and has implications for the validity of approximations such as the instantaneous thermalization of small clusters. Understanding of restructuring for nanosize clusters is not well developed [83, 84]. Without it, clusters would grow into fractals [80, 81] rather than nanocrystals. For larger particles, density measurements and X-ray diffraction data for colloids aggregated from burst-nucleated nanocrystals indicate that they typically have polycrystalline structure. However, their density is close to that of the bulk material [62, 66] implying internal compactification processes. Experimental and indirect modeling evidence suggest [53, 54, 57, 61, 62] that internal restructuring at the constituent nanocrystal contacts and at the aggregate's surface leads to compact polycrystalline particles with smooth surfaces.

The size which separates the two kinetic behaviors in nucleation, $n_c(t)$, is monotonically increasing with time (see Fig. 9). This sharp boundary in the dynamics is an approximation. The short-time particle-size distribution as a function of $n$, depends on the initial conditions. One can generally establish [59, 60, 64] that for large times the size distribution will attain maximum at $n = n_c(t)$. The distribution is thermal for $n < n_c(t)$; Eq. (4). For $n > n_c(t)$, it approaches the shape of a right-side tail of a Gaussian, with the peak of the Gaussian curve, not shown in Fig. 9, located to the left of $n_c$. These properties were derived [59, 60, 64] and also numerically verified by calculating time-dependent distributions for various initial conditions, using a novel efficient numerical integration scheme [59]. Specifically, for large times we have

$$P(n,t) \approx \zeta(t)c_0 \exp\{-[\alpha(t)]^2[n - M(t)]^2\}, \quad \text{for } n > n_c(t). \tag{7}$$

Here the time-dependent quantities are

$$\alpha(t) \approx 1/\sqrt{Zt}\,, \quad M(t) \approx Zt/2\,, \quad \zeta(t) \approx \Omega/\sqrt{Zt}\,, \tag{8}$$

with

$$Z = 64\pi^2 a^3 \sigma c_0 D/(3kT)\,, \tag{9}$$

and $\Omega$ fixed by the initial conditions via the overall normalization of the distribution. Additional mathematical considerations [59] yield $n_c(t) - M(t) \propto \sqrt{t \ln t}$ (with a positive coefficient) for the "peak offset." Since $M(t)$ is linear in time, see Eq. (8), the "offset" is sub-leading, and we obtain the linear form for large times,

$$n_c(t) \approx Zt/2\,. \tag{10}$$

The width of the truncated Gaussian is proportional to $1/\alpha(t) \propto \sqrt{t}$. Therefore, the *relative* width of the distribution decreases according to $\sim 1/\sqrt{t}$. The particle size distribution of the nucleated supercritical particles in burst nucleation can be regarded as narrow not in absolute terms, but only relative to the mean particle size. One can also show [59] that the difference $c(t) - c_0$ approaches zero $\sim 1/\sqrt[3]{t}$.

Numerically, the Gaussian shape offers a good approximation [59] for burst-nucleated particle size distributions also for intermediate times, including the case of the initially seeded distributions. Experimentally, it has been challenging to gather data for nucleated nanocrystals because of their non-spherical shapes and tendency to aggregate. The distribution is usually more evenly two-sided around the peak. The peak is broader than the burst-nucleation prediction, and the final particles in many situations stop growing after a certain time or follow different growth modes and mechanisms. These properties are at odds with the predictions of the simplest burst-nucleation model and can be associated with the breakdown of the assumption of instantaneous thermalization of clusters of all the sizes below the critical and with ignoring other growth mechanisms. The latter include cluster-cluster aggregation and additional effects of a possible monomer detachment, beyond the use, following the ideas of [55], of the prefactor $c(t) - c_0$ in the Smoluchowski rate. Monomers' detachment, competing with their capture, and structure/curvature-related surface free-energy differences between particles, are the ingredients for the process of Ostwald ripening [85].

The structural dynamics of very small clusters is not well studied. Up to sizes tentatively estimated [53, 54, 58, 86-88] to correspond to $n_{th} \approx$ 15–25 "monomers" (atoms, ions, molecules, sub-clusters), they should evolve rapidly enough for the assumption of fast thermalization in burst nucleation to be fully justified. Larger clusters are expected to develop a bulk-like core and their internal restructuring can no longer be regarded as very fast, except perhaps close to their surfaces. For times (and peak sizes) such that $n_c(t) > O(n_{th})$, the nucleation model should be regarded as approximate. Certain modifications have been contemplated [64, 89, 90]. These, however, require introduction of additional kinetic parameters which are not understood as well as those of the basic model.

**<u>Diffusional Growth</u>**

In preceding section we emphasized that burst nucleation can produce particles of narrow size distribution but only of rather small diameters. Subsequent growth is dominated by processes which usually broaden the distribution and push its maximum to larger values. One of such processes is growth by consumption of externally controlled supply of diffusing solute monomers. Here we outline a recently reported [24] novel synthetic procedure to achieve a seeded growth of single crystal nickel nanoparticles (see Fig. 7) in polyol, over an extended range of sizes, driven by diffusional transport of ions supplied by dissolution of dispersed nickel basic carbonate salt. Rather than controlling particle size or shape uniformity, the aim here was to synthesize products with high-quality crystalline-face substrates to be used as cores for shell-core electrocatalysts. As shown in [22], these particles offer improved surface-face morphologies as substrates for electro-catalytic applications.

Experimental data and their model analysis for such nickel particle growth in a range of sizes from 30 to 100 nm were reported [22, 24] under conditions that allow a direct verification of the diffusional transport as the process controlling the growth. Data were obtained for several times of particle growth, and used in a model of diffusional transport of the monomers (nickel ions). Nickel nanocrystals (see Fig. 7) were precipitated by growth on Pt nanoparticle seeds of $1.5 \pm 0.5$ nm in diameter. The latter were prepared by burst nucleation. Growth of nickel particles was driven by an access in the concentration of the monomers over 24 hours. Such a slow particle growth ensures high-quality crystalline faces.

For better understanding of the processes of particle growth, samples were evaluated at different times over 24 hours. Thus, crystal structure analysis indicated nanocrystals with crystallite sizes between 12–20 nm of face-centered cubic (FCC) nickel (JCPS 004-0850) were grown without any preferred orientation, see Fig. 11, indicating that the products remained "highly crystalline" in the sense that for such crystallite sizes, crystal faces of the particles remained ideal for applications. Structural studies suggest that nickel growth was driven by diffusional mechanism through monomer-by-monomer attachment. The system was designed in such a manner that the access Ni-ion concentration was maintained nearly constant by the dissolution of nickel basic carbonate. Those expectations were confirmed by modeling, as outlined below.

We can write rate equations for growth of $s \gg 1$ particles dominated by diffusional capture of monomers as:

$$\frac{dN_s}{dt} = (K_{s-1}\Delta c)N_{s-1} - (K_s \Delta c)N_s , \tag{11}$$

where $K_s$ is the same as in Eq. (5), with $n \to s$; and $s$ denotes the number of monomers in the cluster. Concentration access of monomers in solution, here denoted by $\Delta c$, which is approximately given by $\Delta c(t) = N_1(t) - c_{\text{Ni}}$ as discussed earlier, can be set to the constant difference

$$\Delta c = c_{\text{Ni-carb}} - c_{\text{Ni}} \,. \tag{12}$$

The equilibrium concentrations, denoted $c_{\text{Ni-carb}}$ and $c_{\text{Ni}}$, respectively, for dissolution of nickel basic carbonate and nickel in polyol from the bulk materials under the present experimental conditions, is not known in the literature. As the growth process on average consumes ions from the solution, they are replenished by dissolution of abundant dispersed nickel basic carbonate [24].

Further, for most of the process duration the concentration, the quantity $C(t) = N_1(t)$, cf. Fig. 8 and Eq. (1), is maintained approximately at the value which is the equilibrium concentration for nickel ions from bulk nickel basic carbonate in polyol, $N_1(t) = c_{\text{Ni-carb}}$. This concentration of the dissolved Ni ions was measured at various times and remained in the range $12 \pm 2$ ppm. Model results (see below) suggest that the difference in Eq. (12) is much smaller than each of the two equilibrium concentrations. This proximity of the two equilibrium values, for the growth-driving (dissolving) and growing materials, makes the growth process very slow, which yields highly crystalline particles.

The Smoluchowski rate constant is given here by the expression similar to Eq. (5),

$$K_s = 4\pi R_{\text{particle}} D_{\text{ion}} = A s^{1/3} \,, \tag{13}$$

where the second expression separates out the *s*-dependence via the particle radius which grows $\sim s^{1/3}$, for particles with a large number, *s*, of Ni atoms in them. The coefficient $A$ is evaluated later. The continuous-*s* form of Eq. (11) was analyzed [58] and can be solved provided the second- and higher-degree derivatives in *s* are ignored as contributing only higher-order corrections, as compared to the first-order derivative. The result is a convenient analytical form which, for our case of constant $\Delta c$, can be summarized as

$$N(s,t) = \frac{[s^{2/3} - (2At\Delta c/3)]^{1/2}}{s^{1/3}} N\!\left([s^{2/3} - (2At\Delta c/3)]^{3/2}, 0\right) , \tag{14}$$

where $N(s, 0)$ is the initially seeded particle size distribution, calculated in terms of the effective numbers $s$ of Ni atoms in the volumes of the seed particles. The latter are assumed to be rapidly overgrown by Ni and therefore the growth kinetics is taken identical to that of the seeds being Ni, with the short-time differences ignored. The role of the seeds was only to provide a well-defined initial size distribution.

This expression, Eq. (14), is explained in a schematic in Fig. 12. If the initially seeded, at $t = 0$, size distribution is between $s_{\min}(0)$ and $s_{\max}(0)$, i.e., the function $N(s, 0)$ is practically zero outside the range $s_{\min}(0) < s < s_{\max}(0)$, then the distribution at a later time, $t > 0$, is shifted to the larger range of values, $s_{\min}(t) < s < s_{\max}(t)$,

$$[s_{\min}(t)]^{2/3} = [s_{\min}(0)]^{2/3} + (2A\Delta c/3)t,$$

(15)

$$[s_{\max}(t)]^{2/3} = [s_{\max}(0)]^{2/3} + (2A\Delta c/3)t.$$

In addition to the shift to larger-$s$ values, the shape of the distribution is also changing, and one can show [58] that the size-distribution gradually broadens. The measured particle size distribution at various times [24] is shown in Fig. 13. The average diameter of the particles, plotted in Fig. 14, illustrates their growth. During the first six hours, the distribution remains fairly narrow and symmetrical (see Fig. 13). At later times, it broadens and becomes more skewed towards larger diameters. Figure 14 also shows the half-width of the distribution, calculated as the standard deviation.

The present model does not account for additional kinetic processes and uses a number of mathematical approximations [58]. Otherwise it would yield a further broadening of the calculated distribution in Eq. (14), as well as make it small but nonzero outside the indicated range $s_{\min}(t)$ to $s_{\max}(t)$, even in the case of initial distributions which strictly vanish outside the range $s_{\min}(0)$ to $s_{\max}(0)$. These effects were ignored because the experimental data [24] and knowledge of the various microscopic parameters of the additional processes involved are limited, and the initial distribution, $N(s, 0)$, is also not known exactly. As seen in Fig. 13, the particle distribution is not overly distorted during the observed growth (except, perhaps, for the longest times), and therefore, we can assume that its average growth is well represented by the equal offset, Eq. (15), the same as for the two extreme values in terms of the variable $s^{2/3}$,

$$[s_{average}(t)]^{2/3} \approx [s_{average}(0)]^{2/3} + (2A\Delta c/3)t . \qquad (16)$$

To interpret the data by using Eq. (16), we note that the volume of the primitive unit cell for the FCC Ni (JCPS 004-0850) is $V_0 = 60.70$ Å$^3$. The number of Ni atoms, $s$, in a spherical volume of radius $R_{particle}$, is $s = 4\pi R_{particle}^3/3V_0$. This allows evaluating the experimental values for the average particle size in terms of the number of atoms, $s_{average}(t)$, from their average diameters. The ionic radius for Ni is $a = 0.83$ Å. There is no known estimate of the hydrodynamic radius for diffusion of Ni ions in polyol. Therefore we use $a$ as an approximate value. The diffusion constant $D_{ion}$ is estimated as $kT/6\pi\eta a$, where the viscosity of polyol at our working temperature of $T = 180$ ºC can be estimated [91] from the relation $\eta = \eta_\infty \exp[T_0\delta/(T - T_0)]$, where $\eta_\infty, T_0, \delta$ are given in [91]. For the coefficient $A$ in Eq. (13), we then get $A = (48\pi^2 V_0)^{1/3} D_{ion} = 8.27 \times 10^9$ nm$^3$/sec. The average diameter of the seeds [22], 1.5 nm, corresponds to volume of approximately $s_{average}(0) \approx 29$ unit cells if it were filled with Ni atoms, as explained earlier. With these parameters, Eq. (16) can be used to estimate the effective excess concentration, $\Delta c(t > 0)$, see Eq. (12), plotted in Fig. 15.

Figure 15 further confirms the validity of the assumptions made. The excess concentration difference, $\Delta c$, is much smaller than either one of the two equilibrium concentrations (recall that $c_{Ni}$ is about 12 ppm), and it is approximately constant except for the largest experimental times. For at least the first 6 hours, the supply of the excess Ni ions remained at a constant level due to dissolution of nickel basic carbonate. At the later stages, the particle growth rate somewhat decreases, indicating that this source is being depleted, and the growth process by the present mechanism will ultimately stop.

**Colloid Synthesis Driven by Supply of Nanoparticles**

As they reach sizes up to a couple of 10 nm via burst nucleation and further growth, nanocrystals can in many cases begin to at the same time aggregate, becoming the "monomers" for the formation of polycrystalline colloids. This important two-stage mechanism [1] for synthesis of uniform colloids is shown in Fig. 16. Colloids thus formed have average sizes from a fraction to a couple of microns. Nearly uniform colloid particles of various chemical compositions and shapes have been reported [26, 28, 45, 53, 54, 62, 66, 92-112], with structural properties usually consistent with the two-stage growth mechanism. Typically, spherical colloids were found to have X-ray patterns which are characteristic of highly polycrystalline materials, as measured by peak broadening; these included ZnS [106], CdS [53, 54, 103], $Fe_2O_3$ [104], Au, Ag, and other metals [26, 28, 62, 74, 94, 99, 110]. Many nearly monodispersed inorganic colloids consist of nanocrystalline subunits [26, 28, 45, 53, 54, 62, 66, 92, 94-112], with the sizes of the latter [28, 62, 100] consistent with the dimensions of the precursor nanoparticles formed in solutions. Composite particle structure has also been reported for some uniform non-spherical colloids [45, 92, 104, 105, 108], but these findings are not conclusive enough to confirm the two-stage mechanism.

Here we model the process with simplifications that allow us to avoid introduction of unknown parameters. Improvements that allow a better agreement with experiments are described later. Details can be found in [52-54, 57, 62, 65, 77, 78]. The particles are assumed to primarily grow by irreversible capture of singlets, which is a good approximation for the situation with an already well formed and dominating peak, see Fig. 8. The emergence of the peak is commented on later. We use the by now familiar formulation for singlet capture by the $s \geq 1$ aggregates,

$$\frac{dN_s}{dt} = (K_{s-1}C)N_{s-1} - (K_s C)N_s, \quad \text{for} \quad s \geq 2, \tag{17}$$

$$\frac{dN_2}{dt} = \frac{1}{2}(K_1 C)C - (K_2 C)N_2, \tag{18}$$

$$\frac{dC}{dt} = \rho - \sum_{s=2}^{\infty} s \frac{dN_s}{dt} = \rho - K_1 C^2 - C \sum_{s=2}^{\infty} K_s N_s \,. \tag{19}$$

We use the notation alluded to in Eq. (1) and Fig. 8, with the simplest Smoluchowski rate expression. These assumptions are accepted in the literature [57, 58, 62, 113-115]. Other processes such as cluster-cluster aggregation [80, 81], detachment [61, 62] and exchange of singlets (ripening) [49] also affect the growth and mostly broaden the peak. Some will be addressed later. Regarding the internal (and on-surface) restructuring processes, experimentally one finds [28, 62, 97, 101, 102, 111, 112] that in the two-stage synthesis the growing colloids rapidly restructure to become compact, while remaining polycrystalline, with approximately bulk-like density and typically (though not always) spherical shape and relatively smooth surface. Without such restructuring they would grow fractal [80, 116].

Generally for growth "fed" by the supply of monomers, if the latter are constantly replenished then the size distribution will not develop a peak. It will rather be dominated by a large shoulder at small $s$. If the supply is limited except for the large initial infusion, then only small particles will be formed and there will be no further fast growth. An important finding in colloid synthesis [58, 62] has been that there are protocols of singlet supply, at the rate $\rho(t)$, cf. Eq. (1), which is selected to have a properly decreasing time dependence, that yield size distributions which grow relatively narrow-peaked at large $s$. Furthermore, it turns out that the process of burst-nucleated nanocrystals growing past the nucleation barrier, "feeds" the colloid growth just at such a rate.

Growth of colloids occurs for the appropriate chemical conditions in the system, usually set by the ionic strength and pH. Surface potential should be close to zero, i.e., near the isoelectric point, and/or the electrostatic screening should be substantial enough to avoid electrostatic barriers. These conditions allow fast irreversible nanocrystal attachment [28, 62, 97, 101, 102, 111, 112]. Particles consist of $s$ incorporated nanocrystalline domains, the latter originating from captured nanocrystals and therefore not precisely identical. Without seeding, Eqs. (17)–(19) are solved with the initial conditions $N_{s=1,2,3,\ldots}(0) = 0$, and therefore the time dependence arises entirely from the function $\rho(t)$. For the rate constant we presently use

$$K_s = 4\pi R_p D_p s^{1/3} \,. \tag{20}$$

Here $R_p$ and $D_p$ are the effective primary particle radius and diffusion constant, to be discussed shortly. A numerical calculation result for a model of the type outlined here is shown in Fig. 17. It shows the emergence of the peak and then "size selection" occurring due to the growth process practically freezing even when looked at in 10-fold time increments.

Figure 17 was obtained with the "feed" function, $\rho(t)$, calculated as follows. The rate of production of the supercritical clusters in burst nucleation involves $c(t)$, see Eq. (6). However, relations of the burst-nucleation as the only process cannot be used to calculate the latter. Rather, we use the approximate relation [62],

$$\frac{dc}{dt} = -n_c \rho \,, \tag{21}$$

combined with Eqs. (3), (5) and (6). This yields [62] a closed system of equations for $c(t)$, presented shortly. As the burst-nucleated, growing supercritical nanoparticles are additionally consumed by the secondary aggregation, the solute species of supersaturated concentration $c(t)$ are dynamically rebalanced to be partially incorporated in subcritical embryos, as well as in the supercritical nanoparticles which in turn are captured into the forming secondary colloids. Equation (20) is an approximation [62] that offers tractability, by ignoring the possible rebalancing of the "recoverable" stored solute species in various part of the particle distributions. It focuses on the loss of the solute-species availability due to the mostly unrecoverable storage in the supercritical nanoparticles and their colloid aggregates. The approximation also ignores direct capture by and detachment from the larger particles (colloids).

The set of equations used to calculate $\rho(t)$ was obtained [62], to be solved with the initial condition provided by the supersaturation $c(0) \gg c_0$, if we assume instantaneous (very fast) production of the large supersaturated solute species concentration to initiate the process,

$$\frac{dc}{dt} = -\frac{2^{14}\pi^5 a^9 \sigma^4 D_a c^2}{(3kT)^4[\ln(c/c_0)]^4} \exp\left\{-\frac{2^8\pi^3 a^6 \sigma^3}{(3kT)^3[\ln(c/c_0)]^2}\right\}, \tag{22}$$

$$\rho = \frac{2^5\pi^2 a^3 \sigma D_a c^2}{3kT\ln(c/c_0)} \exp\left\{-\frac{2^8\pi^3 a^6 \sigma^3}{(3kT)^3[\ln(c/c_0)]^2}\right\}, \tag{23}$$

where $D_a$ denotes the diffusion constant of the solutes, and other notation was defined earlier. We next further comment on the model and it's approximations and assumptions, possible improvements of which are addressed in the next section. In fact, Fig. 17 was based on one such improved variant of the model for Au colloids [57].

The Smoluchowski rate expression for the irreversible-aggregation rate constant on encounters of diffusing particles of sizes $s_1$ and $s_2$, multiplying the product of their concentrations in the rate expression, can be written as

$$K_{s_1,s_2 \to s_1+s_2} = 4\pi\left[R_p\left(s_1^{1/3} + s_2^{1/3}\right)\right]\left[D_p\left(s_1^{-1/3} + s_2^{-1/3}\right)\right], \tag{24}$$

for sizes which are not exactly equal. For the singlet capture case considered before we took the limiting form for $s_1 = s \gg 1$, and $s_2 = 1$. However, when both $s$ values are small, nontrivial changes were involved in the latter approximation. Equation (24) also assumes that the diffusion constant of $s$-singlet, dense particles is inversely proportional to the radius, i.e., to $s^{-1/3}$, which might not be accurate for very small, few-singlet aggregates. Furthermore, the radius of a representative $s$-singlet, dense colloid is estimated as $R_p s^{1/3}$. Nanoparticles actually have a distribution of radii. However, since nanoparticle capture rate by the aggregates is approximately proportional to their radius times their diffusion constant, this rate will not be that sensitive to each specific particle's size, because its diffusion constant is inversely proportional to its radius. Thus, the product is well approximated by using a single typical value, $R_p$.

However, the estimation of $R_p$ requires some discussion. Indeed, nanoparticles are not necessarily captured immediately after growing past the nucleation barrier as was assumed in writing our simplest rate expression for their availability. Their coarsening after nucleation but before capture can be approximately accounted for by using the experimentally determined typical nanosize singlet linear dimension, $2R_{exp}$, instead of attempting to calculate the size distribution dynamically as a function of time. The representative radius of the *s*-singlet particle in the first factor in Eq. (24), and generally entering as $R_p s^{1/3}$ in rate expressions, was thus estimated by replacing it with

$$R_p s^{1/3} = 1.2 R_{exp} s^{1/3} \; . \tag{25}$$

Here the factor $(0.58)^{-1/3} \approx 1.2$ derives from the filling factor, 0.58, of a random close-packing of spheres [117], and allows to approximately account for that as the growing colloid particle compactifies by restructuring, not all its volume will be crystalline. A fraction will consists of amorphous "bridging regions" between the nanocrystalline subunits.

The approximations described above can lead to nonconservation of the total amount of matter. This can be corrected [62] by normalizing the calculated final size-distributions to have the product particles contain the correct amount of matter as initially supplied. This seems not to play a significant role in the dynamics. Some additional model details are elaborated on the literature [52-54, 56, 57, 62, 65].

**Discussion and Summary**

The formulated model of colloid synthesis was applied for a semi-quantitative description (without adjustable parameters) of the processes of formation of spherical colloids of metals Au [52, 54, 56, 57, 62, 65], Ag [52, 65], and generally metals [77, 78], a salt CdS [53, 54], as well as to qualitatively explain the synthesis of microspheres of an organic colloid, Insulin [93]. There have been studies aiming at improving the model for better quantitative agreement with experimental results for CdS [53, 54], Au, and Ag [52, 65]. For spherical CdS, colloid radius distribution was measured at several times during the process and for varying protocols of releasing the solutes. When solute ions (or atoms/molecules) are not released as a batch or externally supplied, we have to add to the model equations the rate terms for their production in chemical reactions. For many common colloid synthesis situation, the experimental identification and even more so modeling of the chemical kinetics of the relevant solute species are not well researched.

Numerical simulations yield useful information on the control of the growth process. Specifically, it was found that the parameters of the nanocrystal nucleation, specifically, the effective surface tension, $\sigma$, and equilibrium concentration, $c_0$, affect the time scales of the onset of "freezing" of the secondary aggregation. The parameters of the secondary process, which are addressed in more detail shortly, were found to set the size of the final products. Generally, colloid sizes obtained with the "minimal" model [52-54, 56, 57, 62, 65] were of the correct order of magnitude, but consistently smaller than the experimentally observed values. It seems that the model gives too many secondary particles, which then on average grow to sizes smaller than those measured.

Improving the model has thus aimed at revisiting the aggregation assumptions. One could argue [57, 65], for instance, that for the smallest aggregates, those consisting of few particles, the diffusional expressions for the rates, which are anyway ambiguous for small clusters as discussed in connection with Eq. (24), should be modified. In order to avoid introduction of many adjustable parameters, the processes of irreversible monomer capture were still considered to be the dominant in that only terms with rates $K_{s\geq 1, 1 \to s+1}$ were kept. However, the rate $K_{1,1\to 2}$ was multiplied by a "bottleneck" factor, $f \ll 1$. This modification attempts to account for that the pairwise merging of two singlets (and, in fact, of other very small aggregates) may require a substantial restructuring. This reduces the rate of a successful formation of a bi-crystalline entity. Indeed, the two nanocrystals may instead diffuse apart, or merge into a single larger nanocrystal, the latter process effectively contributing to the rate $K_{1,1\to 1}$ of a process not included in the model. Data fits [54, 57, 65] yield $f$ values of order $10^{-3}$ or smaller.

A different possible starting point for improving the simplest model [53, 54] is the observation that this model already assumes a "bottleneck" for particle aggregation: only the singlet-capture is considered, whereas the processes with both $s_1 > 1$ and $s_2 > 1$ are ignored.

This was originally motivated by the observation that colloid-size particles were never experimentally seen to pair-wise merge in solution. Apparently, various internal restructuring processes that cause compactification of the growing colloids and mediate the incorporation of the constituent nanoparticles, are not effective at incorporating larger aggregates. Indeed, even the incorporated single nanoparticles mostly retain their crystalline cores to yield the final polycrystalline colloids.

However, the aforementioned experimental evidence only applied to larger colloids/aggregates. The improved model allows sufficiently small clusters with $s_{\max} \geq s \geq 1$, up to a certain number of nanoparticle domains in them, $s_{\max} > 1$, to also be rapidly incorporated into the aggregates. Thus, in addition to the monomer-cluster aggregation, the model [53, 54] includes cluster-cluster (i.e., $s_{1,2} \geq 1$) aggregation with rates given by Eq. (24), but only as long as at least one of the sizes, $s_{1,2}$, does not exceed $s_{\max}$. This sharp cutoff is obviously an approximation, but it offers the convenience of a single new adjustable parameter, $s_{\max}$. Data fits for CdS and Au spherical particles have yielded quantitative agreement with experiments, with values of $s_{max}$ ranging [53, 54] from ~15 for Au, to ~25 for CdS. These values are reasonable as defining "small" aggregates and consistent with a similar concept of the cluster size estimate, $n_{th}$, discussed in the concluding paragraph of the section on burst nucleation, beyond which size the atomistic clusters develop a "bulk-like" core. Indeed, a numerical estimate for AgBr nanoaggregates in solution [88] suggests that $n_{th}$ is comparable to or somewhat larger than ~18. Another appealing feature of this model is that the added cluster-cluster aggregation at small sizes offers a mechanism for the formation of the initial peak in the secondary-particle distribution.

Finally, we note that the simplest model and improved ones (with more parameters), all require substantial computational resources. Simulation speed-up techniques, not reviewed here, for the kinetics of larger clusters have been reported and utilized [52-54, 58, 65].

In summary, we considered examples of chemical methods used to synthesize highly dispersed metallic particles with controlled properties and outlined their practical importance in industrial applications. The challenges involved in developing new synthetic procedures that yield materials meeting the demands of specific technologies were also reviewed. Models of particle growth processes that offer a qualitative or even semi-quantitative understanding of the mechanisms of particle formation and size control were detailed. Recent results for shape selection were also referenced but the mechanisms involved were not discussed in detail. The preparation protocols and materials surveyed here, including on-surface nanostructure growth, have the potential to bridge key scientific sub-disciplines and result in the incorporation of electrochemically controlled processes in the design and manufacturing of materials for advanced technologies.

Indeed, the field is in dire need for additional research. Theoretical understanding is presently at the stage when we have tentatively identified certain specific mechanisms and conditions which seem to control the growth of crystalline or polycrystalline particles of relatively uniform size and/or shape, the former from nanosizes (nanoparticles) to order of micron sizes (colloids). Experimental data are limited to the observation of the final products. Experimental results for time dependent, kinetic processes, as well as detailed morphological data are scarce and getting them systematically for a wide range of representative systems would benefit future model development.

The authors thank P. Atanassov, V. Gorshkov, I. Halaciuga, D. Robb, and O. Zavalov for rewarding scientific interactions and collaboration, and acknowledge research support by the US ARO under Grant W911NF-05-1-0339.

**References**

1. Hartl K, Mayrhofer K, Lopez M, Goia D, Arenz M (2010) Electrochem Commun 12:1487–1489
2. Kinnunen NM, Suvanto M, Moreno MA, Savimaki A, Kallinen K, Kinnunen T-JJ, Pakkanen TA (2009) Appl Catal, A 370:78–87
3. Mbang T, Ketcha J, Geron C (2005) Ann Chim Sci Mat 30:149–170
4. Zhang J, Lima FHB, Shao MH, Sasaki K, Wang JX, Hanson J, Adzic RR (2005) J Phys Chem B 109:22701–22704
5. Kokkinidis G, Stoychev D, Lazarov V, Papoutsis A, Milchev A (2001) J Electroanal Chem 511:20–30
6. Balantrapu K, Goia DV (2009) J Mater Res 24:2828–2836
7. Jitianu A, Kim MS, Andreescu D, Goia DV (2009) J Nanosci Nanotechnol 9:1891–1896
8. Toschev S, Milchev A, Popova K, Markov I (1969) Compt Rend Acad Bulg Sci 22:1413–1416
9. Pillai S, Catchpole KR, Trupke T, Green MA (2007) J Appl Phys 101: 093105, 8 pages
10. Morrow BJ, Matijevic E, Goia DV (2009) J Colloid Interface Sci 335:62–69
11. Andreescu D, Sau TK, Goia DV (2006) J Colloid Interface Sci 298:742–751
12. Matijevic E (2011) Fine particles in medicine and pharmacy, Springer, New York
13. Kwon JH, An JY, Jang H, Choi S, Chung DS, Lee MJ, Cha HJ, Park JH, Park CE, Kim YH (2011) J Polym Sci, Part A: Polym Chem 49:1119–1128
14. Nam HJ, Jung DY, Park YK, Park S (2010) Appl Surf Sci 256:2066–2072
15. Sabat RG, Santos MJL, Rochon P (2010) Int J Photoenergy, 2010:698718, 5 pages
16. Goia DV (2004) J Mater Chem 14:451–458
17. Suganuma K, Wakuda D, Hatamura M, Kim KS (2007) In: HDP Proc, pages 1–4, Shanghai
18. Zarbov M, Brandon D, Gal-Or L, Cohen N (2006) In: Boccaccini A, VanderBiest O, Clasen R (ed) Electrophoretic deposition: Fundamentals and applications 314:95–100
19. Koga K, Matsunaga T, Nakamura WM, Nakahara K, Kawashima Y, Uchida G, Kamataki K, Itagaki N, Shiratani M (2011) Thin Solid Films 519:6896–6898
20. Park JH, Song J, Lee JH, Lee JC (2012) J Kor Phys Soc 60:2054–2057
21. Temple TL, Bagnall DM (2011) J Appl Phys 109:084343), 13 pages

22. Goia DV, Lopez M, Sevonkaev I (2012) Method of manufacture of size-controlled metal particles by seeding method. Patent US 2012 / 0238443 A1
23. Halaciuga I, LaPlante S, Goia DV (2009) J Mater Res 24:3237–3240
24. Sevonkaev I, Privman V, Goia D (2012) Growth of highly crystalline nickel particles by diffusional capture of atoms (submitted)
25. Farrell BP (2010) Investigation of shape anisotropy in precipitated silver and silver/palladium particles. Ph.D. thesis, Clarkson University, Potsdam, NY, USA
26. Halaciuga I, Goia DV (2008) J Mater Res 23(06):1776–1784
27. Sevonkaev I, Farrell B, Goia D Preparation of spherical platinum particles with controlled size and internal structure (submitted)
28. Goia D, Matijevic E (1999) Colloids Surf , A 146:139–152
29. Grass RN, Albrecht TF, Krumeich F, Stark WJ (2007) J Mater Chem 17:1485–1490
30. Hunt E, Pantoya M (2005) J Appl Phys 98:034909, 8 pages
31. Ionkin AS, Fish BM, Li ZR, Lewittes M, Soper PD, Pepin JG, Carroll AF (2011) ACS Appl Mater Interfaces 3:606–611
32. Shi Y, Liu J, Yan Y, Xia Z, Lei Y, Guo F, Li X (2008) J Electron Mater 37:507–514
33. Vasylkiv O, Sakka Y, Skorokhod VV (2006) J Nanosci Nanotechnol 6:1625–1631
34. Fu W, Yang H, Chang L, Li M, Bala H, Yu Q, Zou G (2005) Colloids Surf , A 262:71–75
35. Roy R, Njagi J, Farrell B, Halaciuga I, Lopez M, Goia D (2012) J Colloid Interface Sci 369:91–95
36. Halaciuga I, Njagi JI, Redford K, Goia DV (2012) J Colloid Interface Sci 383:215–221
37. Farrell BP, Andreescu D, Eastman CM, Goia D (2005) CARTS Proc Palm Springs, California March 21-24, ECIA, Alpharetta, GA
38. Mehta M, Burn I, Basak S, Goia D, Soni H, Spang D (1999) CARTS Proc New Orleans, Louisiana March 15–19, ECIA, Alpharetta, GA
39. Schuldiner S, Rosen M, Flinn DR (1970) J Electrochem Soc 117:1251–1259
40. Kim S, Kim SK, Park S (2009) J Am Chem Soc 131:8380–8381
41. Sonnichsen C, Franzl T, Wilk T, von Plessen G, Feldmann J, Wilson O, Mulvaney P (2002) Phys Rev Lett 88:0774024, 4 pages
42. Yu S, Colfen H, Mastai Y (2004) J Nanosci Nanotechnol 4:291–298
43. Farrell BP, Lu L, Goia DV (2012) J Colloid Interface Sci 376:62–66

44. Sevonkaev I, Privman V (2009) World J Eng 6:P909, 2 pages
45. Sevonkaev I, Goia DV, Matijevic E (2008) J Colloid Interface Sci 317(1):130–136
46. Sevonkaev I (2009) Size and shape of uniform particles precipitated in homogeneous solutions. Ph.D. Thesis, Clarkson University, Potsdam, NY, USA
47. Milchev A (2008) Russ J Electrochem 44:619–645
48. Milchev A, Kruijt W, Sluyters-Rehbach M, Sluyters J (1993) J Electroanal Chem 362:21–31
49. Gorshkov V, Privman V (2010) Physica E 43:1–12
50. Gorshkov V, Zavalov A, Privman V (2009) Langmuir 25:7940–7953
51. Gorshkov V, Zavalov O, Atanassov PB, Privman V (2011) Langmuir 27:8554–8561
52. Halaciuga I, Robb D, Privman V, Goia D (2009) In: Avis D, Kollmitzer C, Privman V (ed) Proc Conf ICQNM 2009, IEEE Comp Soc Conf Publ Serv, Los Alamitos, California, pages 141–146
53. Libert S, Gorshkov V, Goia D, Matijevic E, Privman V (2003) Langmuir 19:10679–10683
54. Libert S, Gorshkov V, Privman V, Goia D, Matijevic E (2003) Adv Colloid Interface Sci 100:169–183
55. Mozyrsky D, Privman V (1999) J Chem Phys 110:9254–9258
56. Park J, Privman V (2000) Recent Res Dev Stat Phys 1:1–17
57. Park J, Privman V, Matijevic E (2001) J Phys Chem B 105:11630–11635
58. Privman V (2002) Mater Res Soc Symp Proc 703:577–585
59. Privman V (2008) J Optoelectron Adv Mater 10:2827–2839
60. Privman V (2009) Ann NY Acad Sci 1161:508–525
61. Privman V, Park J (2001) Mat Processing Proc 1:141–147
62. Privman V, Goia D, Park J, Matijevic E (1999) J Colloid Interface Sci 213:36–45
63. Privman V, Gorshkov V, Zavalov O (2011) In: Sadhal SS (ed) Proc Conf ITP 2011, University of Southern California Press, Los Angeles, California, article 11, pages 4-2–4-12
64. Robb D, Privman V (2008) Langmuir 24:26–35
65. Robb DT, Halaciuga I, Privman V, Goia DV (2008) J Chem Phys 129:184705, 11 pages
66. Matijevic E (1993) Chem Mater 5:412–426
67. Matijevic E, Goia D (2007) Croat Chem Acta 80:485–491

68. Leon-Velazquez M, Irizarry R, Castro-Rosario M (2010) J Phys Chem C 114(13):5839–5849
69. Eastoe J, Hollamby MJ, Hudson L (2006) Adv Colloid Interface Sci 128-130(0):5–15
70. Fendler JH and Tian Y (2007) In: Fendler JH (ed) Nanoparticles and nanostructured films: preparation, characterization and applications, Wiley-VCH Verlag GmbH, Weinheim, Germany, chapter 18, pages 429–461
71. Xia Y, Xiong Y, Lim B, Skrabalak SE (2009) Angew Chem, Int Ed 48:60–103
72. Schmid G (1992) Chem Rev 92:1709–1727
73. Teranishi T, Hosoe M, Tanaka T, Miyake M (1999) J Phys Chem B 103:3818–3827
74. LaMer VK (1952) Ind Eng Chem 44:1270–1277
75. LaMer VK, Dinegar RH (1950) J Am Chem Soc 72:4847–4854
76. Chernov S, Fedorov Y, Zakharov V (1993) J Phys Chem Solids 54:963–966
77. Irizarry R (2010) Ind Eng Chem Res 49:5588–5602
78. Irizarry R, Burwell L, Leon-Velazquez MS (2011) Ind Eng Chem Res 50:8023–8033
79. Nanev CN, Hodzhaoglu FV, Dimitrov IL (2011) Cryst Growth Des 11:196–202
80. Family F, Vicsek T (1991) Dynamics of fractal surfaces. World Scientific, Singapore
81. Godreche C (1992) Solids far from equilibrium. Cambridge University Press, Cambridge
82. Sevonkaev I (2009) (unpublished)
83. Baletto F, Ferrando R (2005) Rev Mod Phys 77:371–423
84. Lewis LJ, Jensen P, Barrat JL (1997) Phys Rev B: Condens Matter 56:2248–2257
85. Voorhees P (1985) J Stat Phys 38:231–252
86. Goia D (1999) (unpublished)
87. Kelton KF, Greer AL (1988) Phys Rev B: Condens Matter 38:10089–10092
88. Shore JD, Perchak D, Shnidman Y (2000) J Chem Phys 113:6276–6284
89. Kelton KF, Greer AL, Thompson CV (1983) J Chem Phys 79:6261–6276
90. Ludwig FP, Schmelzer J (1996) J Colloid Interface Sci 181:503–510
91. Jadzyn J, Czechowski G, Stefaniak T (2002) J Chem Eng Data 47:978–979
92. Bailey JK, Brinker C, Mecartney ML (1993) J Colloid Interface Sci 157:1–13
93. Bromberg L, Rashba-Step J, Scott T (2005) Biophys J 89:3424–3433
94. Crnjak OZ, Matijevic E, Goia DV (2003) Colloid Polym Sci 281:754–759
95. Edelson LH, Glaeser AM (1988) J Am Ceram Soc 71:225–235

96. Goia C, Matijevic E (1998) J Colloid Interface Sci 206:583–591
97. Haruta M, Delmon B (1986) J Chim Phys 83:859–868
98. Hsu WP, Ronnquist L, Matijevic E (1988) Langmuir 4:31–37
99. Jitianu M, Goia DV (2007) J Colloid Interface Sci 309:78–85
100. Lee SH, Her YS, Matijevic E (1997) J Colloid Interface Sci 186:193–202
101. Matijevic E (1985) Ann Rev Mater Sci 15:483–516
102. Matijevic E (1994) Langmuir 10:8–16
103. Matijevic E, Murphy-Wilhelmy D (1982) J Colloid Interface Sci 86:476–484
104. Matijevic E, Scheiner P (1978) J Colloid Interface Sci 63:509–524
105. Morales M, Gonzalez-Carreno T, Serna C (1992) J Mater Res 7:2538–2545
106. Murphy-Wilhelmy D, Matijevic E (1984) J Chem Soc, Faraday Trans 80:563–570
107. Ocana M, Matijevic E (1990) J Mater Res 5:1083–1091
108. Ocana M, Morales MP, Serna CJ (1995) J Colloid Interface Sci 171:85–91
109. Ocana M, Serna CJ, Matijevic E (1995) Colloid Polym Sci 273:681–686
110. Sondi I, Goia DV, Matijevic E (2003) J Colloid Interface Sci 260:75–81
111. Sugimoto T (1987) Adv Colloid Interface Sci 28:65–108
112. Sugimoto T (1992) J Colloid Interface Sci 150:208–225
113. Brilliantov N, Krapivsky P (1991) J Phys A: Math Gen 24:4789–4803
114. Dirksen JA, Benjelloun S, Ring TA (1990) Colloid Polym Sci 268:864–876
115. van Dongen PGJ, Ernst MH (1984) J Stat Phys 37:301–324
116. Schaefer DW, Martin JE, Wiltzius P, Cannell DS (1984) Phys Rev Lett 52:2371–2374
117. German RM (1989) Particle packing characteristics. Metal Powder Industries Federation. Princeton, NJ

**FIGURES**

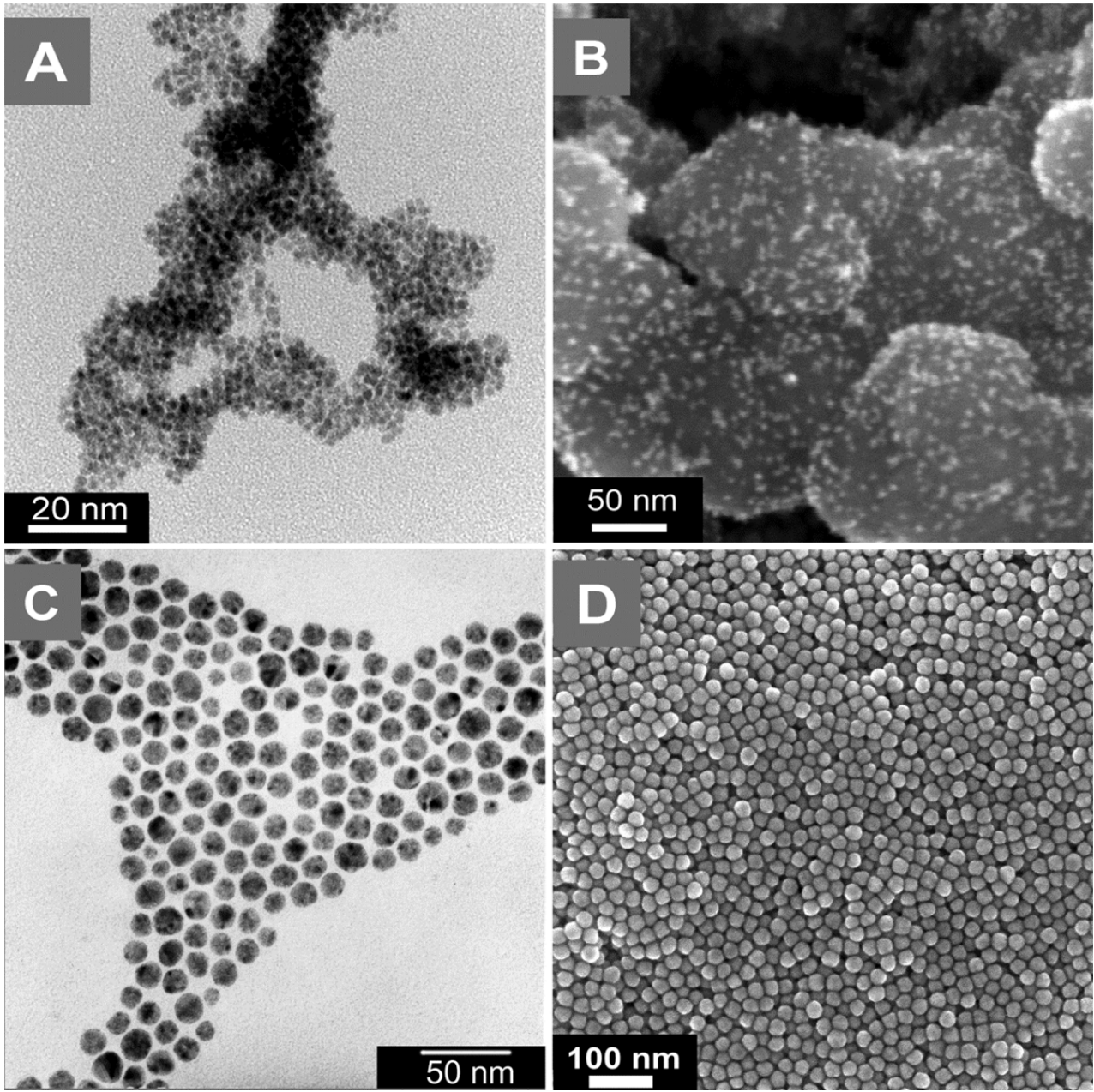

**Figure 1:** (A) Dispersed Pt nanoparticles (~3 nm); (B) the same Pt particles deposited on a carbon substrate; (C) uniform Ag nanoparticles (~ 10 nm) [6, 7]; (D) highly dispersible uniform gold nanoparticles (~ 20 nm).

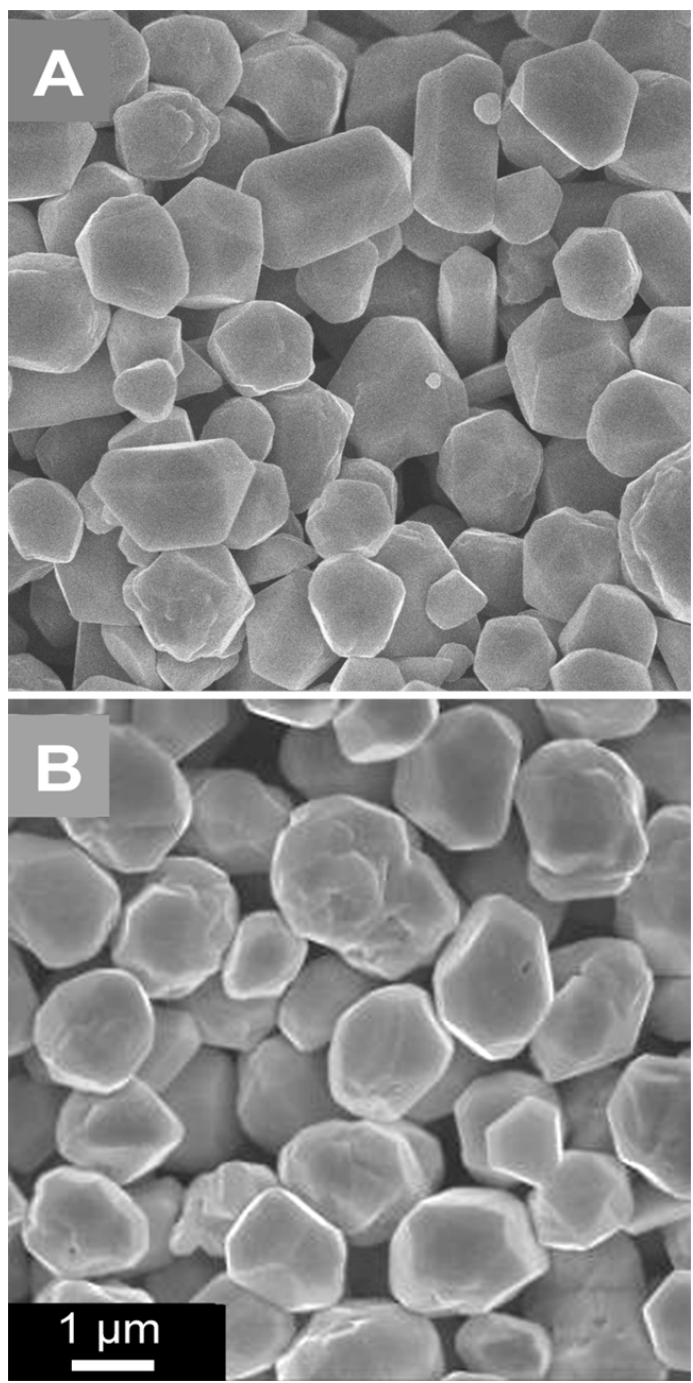

**Figures 2:** (A) Copper crystalline particles obtained by reducing CuCl with ferrous citrate [23]; (B) silver crystalline particles obtained by reducing silver nitrate with ascorbic acid under acidic conditions [25].

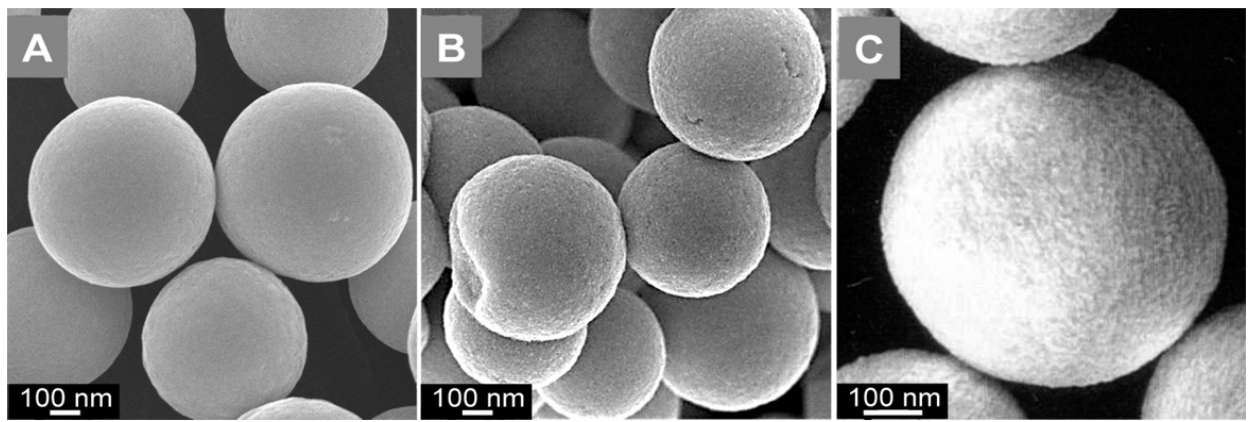

**Figure. 3:** (A) Spherical polycrystalline Ag particles obtained by the reduction of the Ag-TETA complex with ascorbic acid [26]; (B) Pt particles obtained by the reduction of $[Pt(NH_3)_6]^{4+}$ with ascorbic acid [27]; (B) Au polycrystalline particles obtained by the reduction of $HAuCl_4$ with ascorbic acid [28].

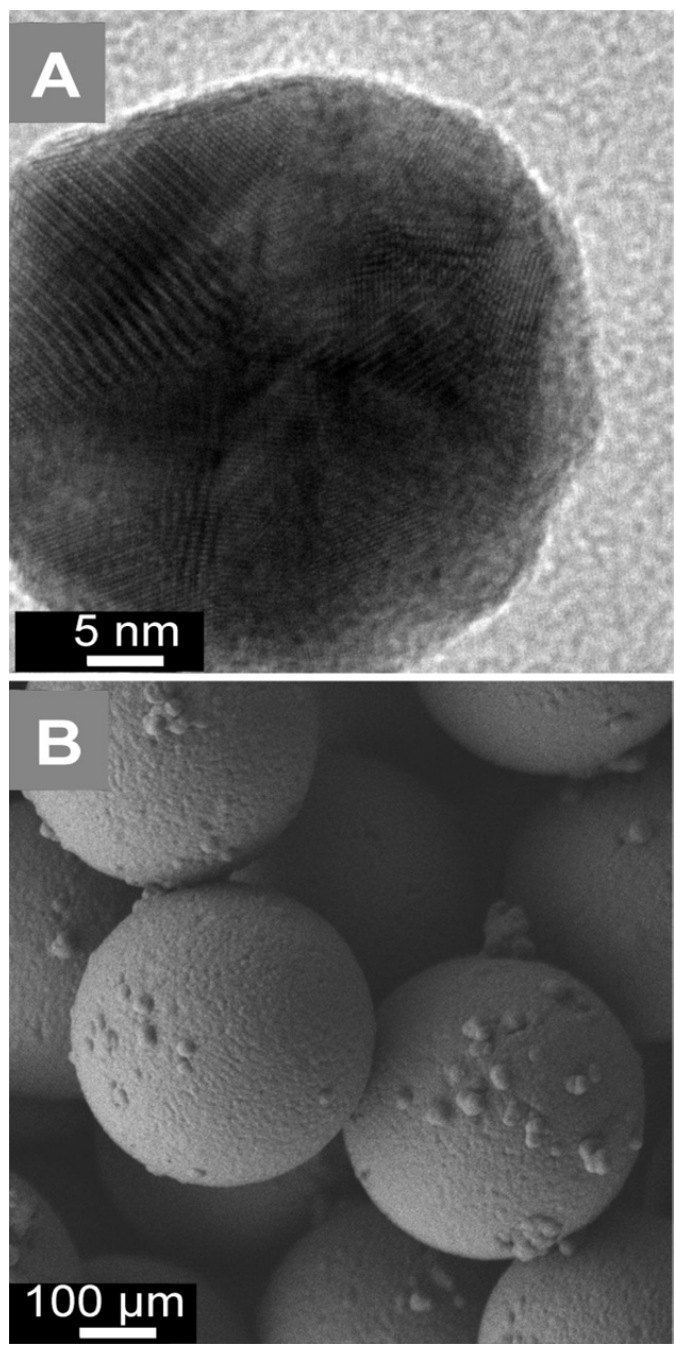

**Figure 4:** (A) Gold nanocrystal covered with a continuous crystalline Pt shell consisting of ~ 5 atomic layers [35]; (B) polymer spheres (4 µm average diameter) coated with a continuous polycrystalline Ni shell; the thickness of the nickel layer is ~ 120 nm and the size of the constituent crystallites is ~ 18 nm [36].

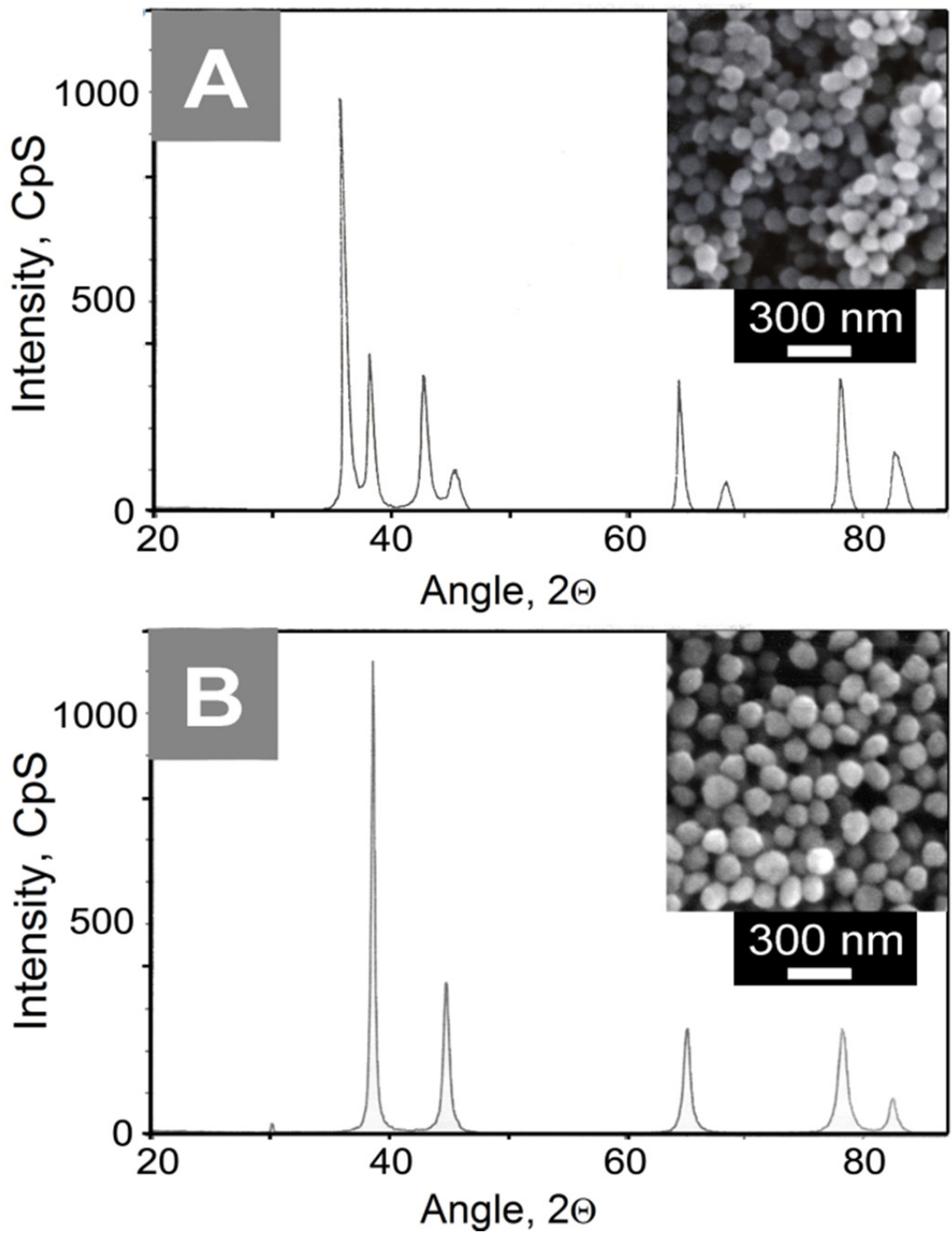

**Figure 5**: (A) XRD of the Ag/Pd core/shell particles demonstrates the peaks of individual metals; (B) XRD shows individual peaks that correspond to AgPd alloy particles. SEMs shown in the insets indicate particles of ~ 120 nm in both cases [37, 38].

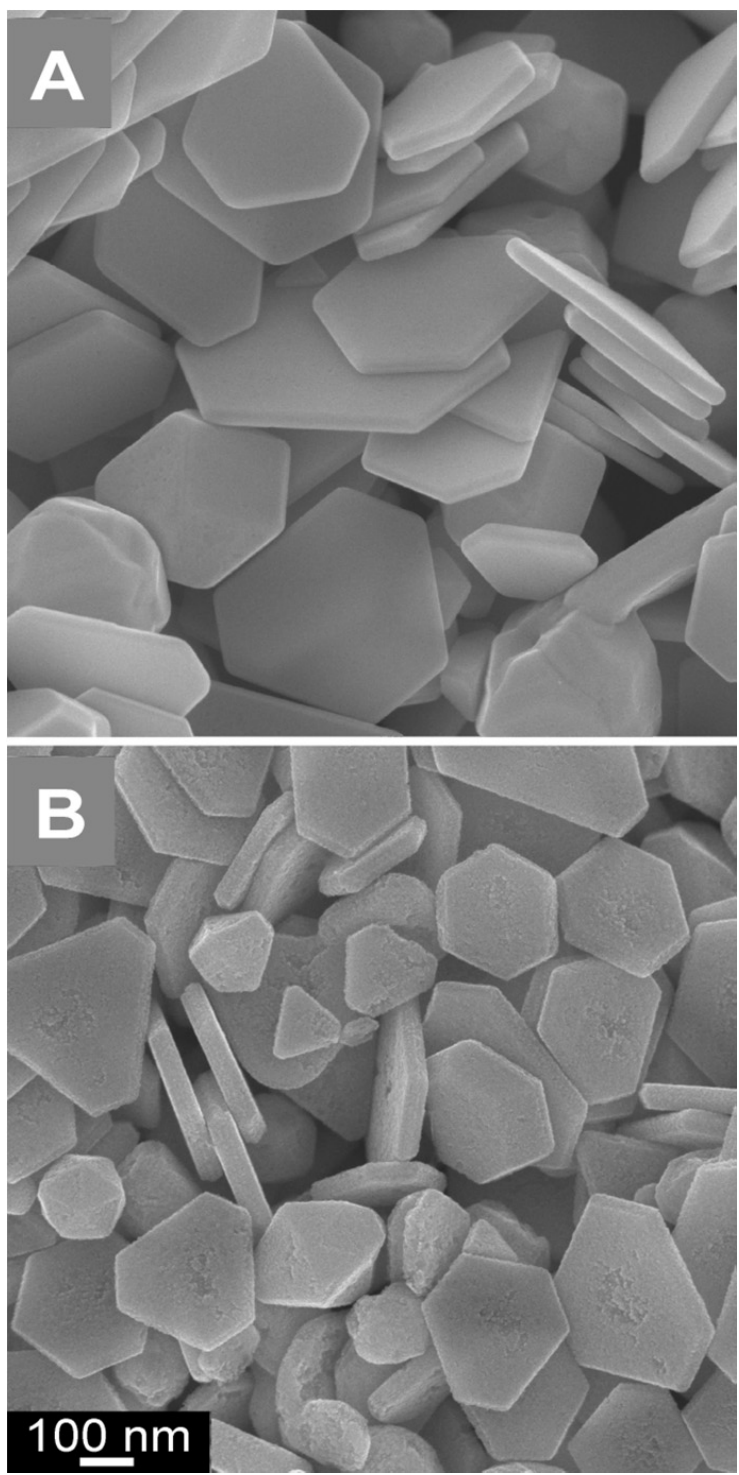

**Figure 6:** (A) Ag nanoplatelets, and (B) AgPd nanoplatelets obtained by controlled nucleation and diffusional growth in acidic solutions [25, 43].

– 39 –

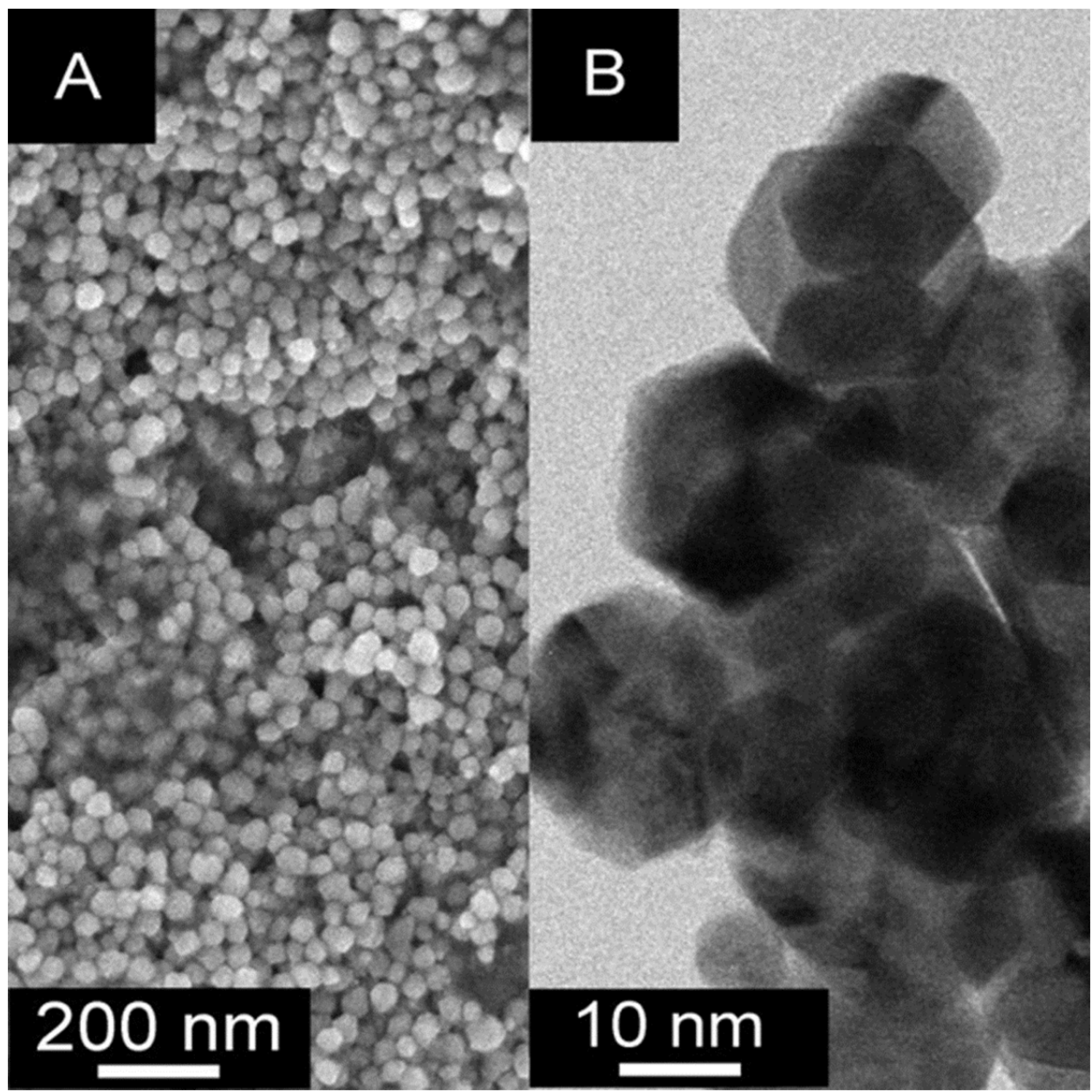

**Figure 7:** (A) FESEM and (B) HRTEM of crystalline Ni particles, obtained by seeded diffusional growth [22, 24].

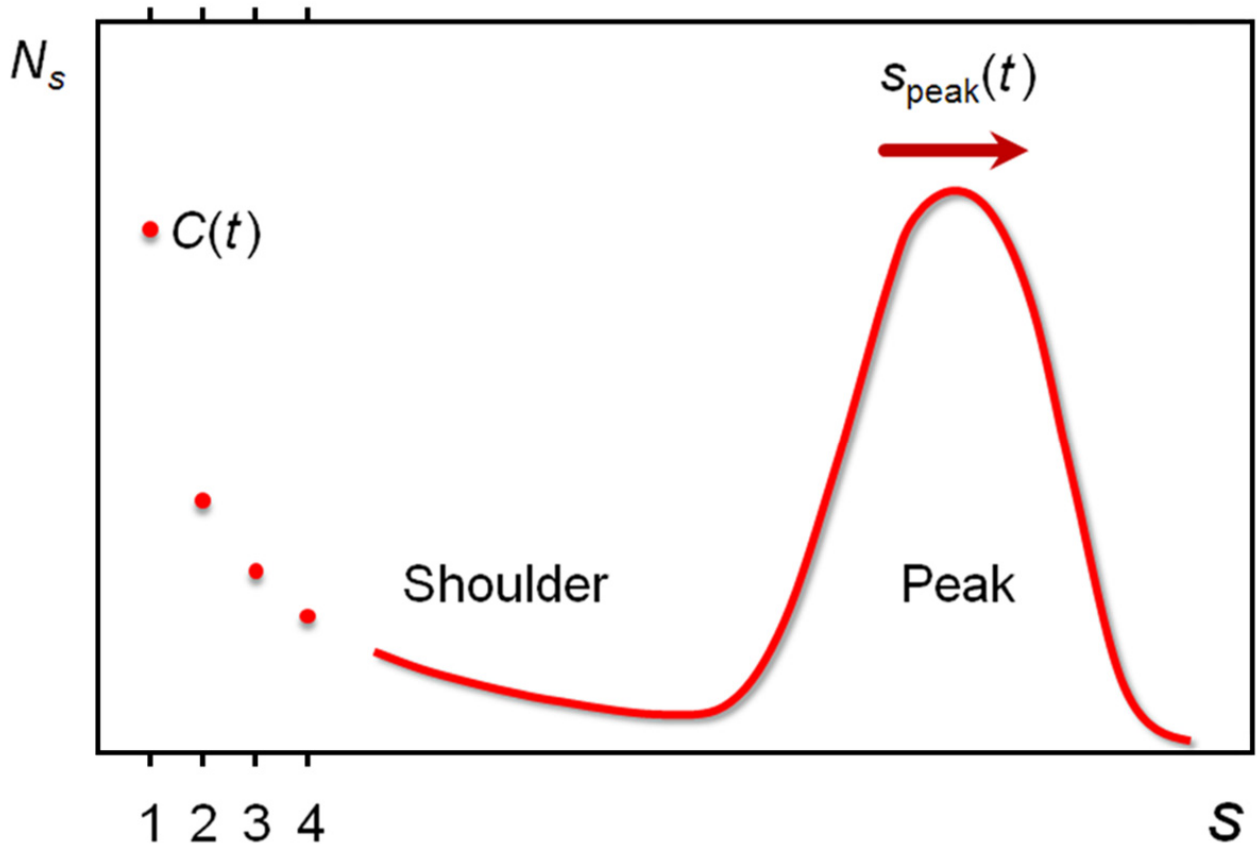

**Figure 8:** The desirable size distribution, $N_s(t)$, with the peak at $s_{peak}(t)$, growing with time but remaining narrow. Particle growth could be controlled by chemically released or nucleated "singlets" with concentration $C(t) = N_1(t)$. Distinct values shown for sizes $s = 1, 2, …$, signify that $s$ is actually discrete, even though for large enough $s$, the function $N_s(t)$ is treated as continuous (in $0 \leq s < \infty$); see text for additional discussion.

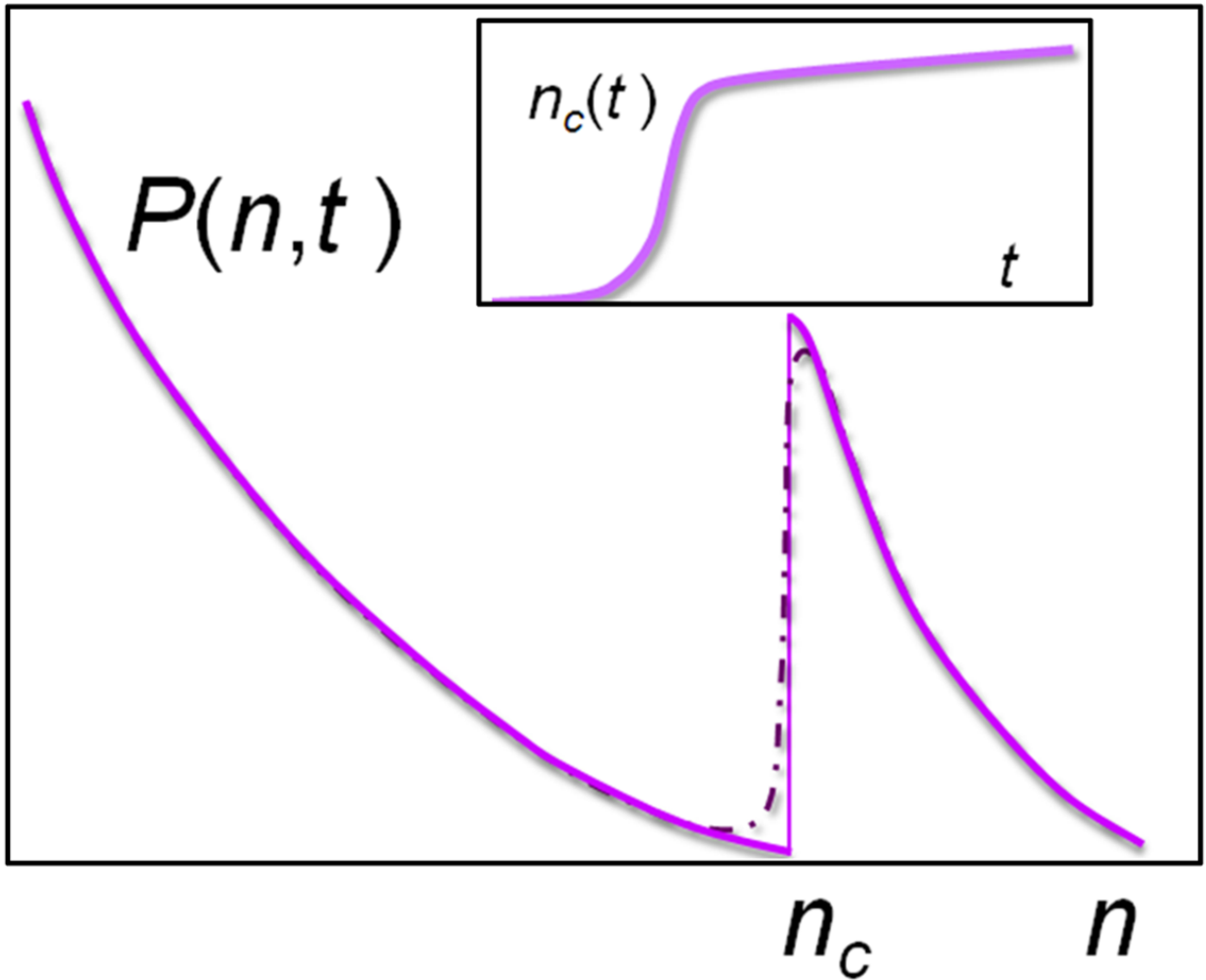

**Figure 9:** Size distribution for burst nucleation at a fixed time. The solid line sketches the approximation described in the text. The actual distribution, shown by the dotted-dashed line, will be steep but smooth at $n_c$. The time dependence of $n_c$ is shown in the inset, including the initial "induction" period, then the "burst," and finally the large-time very slow (small slope) linear growth.

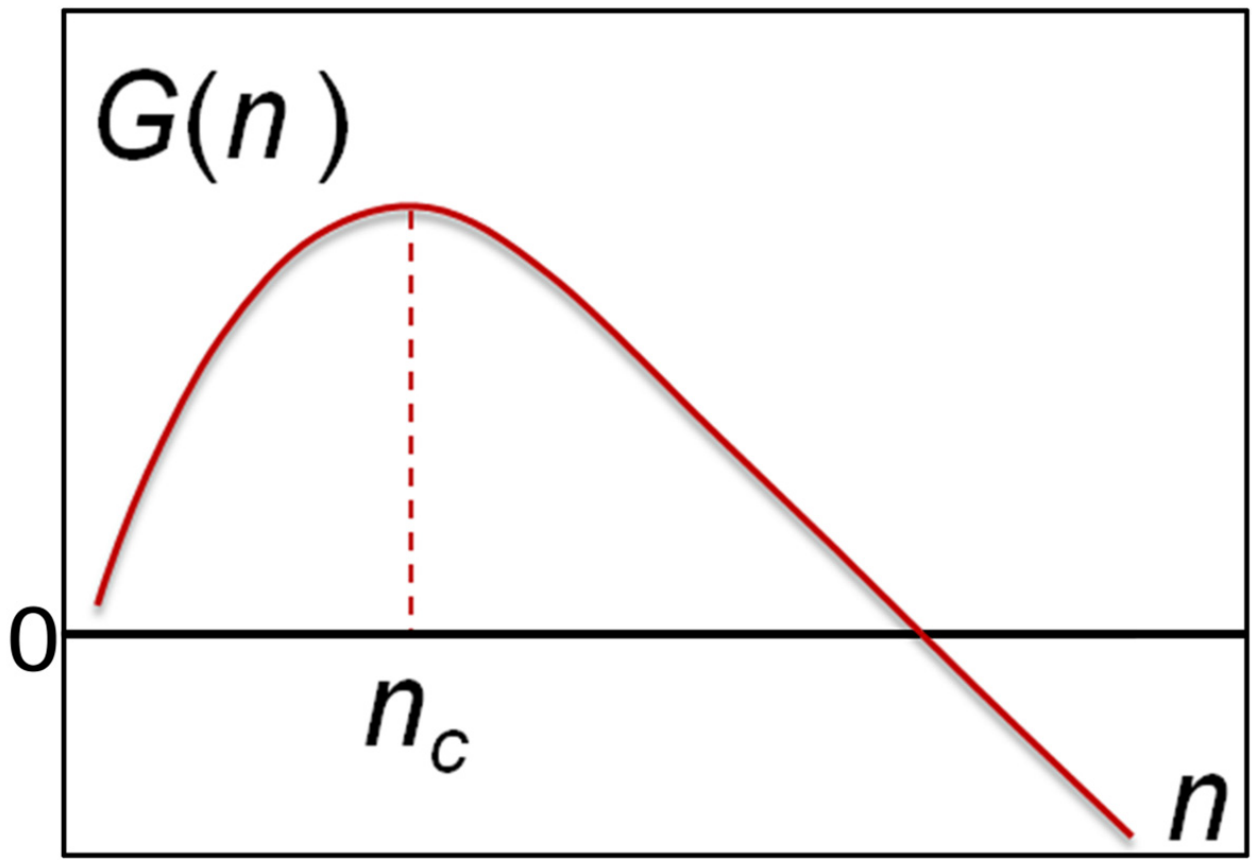

**Figure 10:** Sketch of the free-energy function in Eq. (2). Nucleation approach assumes that up to the barrier, peaked at $n_c$, subcritical clusters are thermally distributed. Supercritical clusters grow irreversibly. Thus the size-distribution of the latter is not controlled by the shown free-energy.

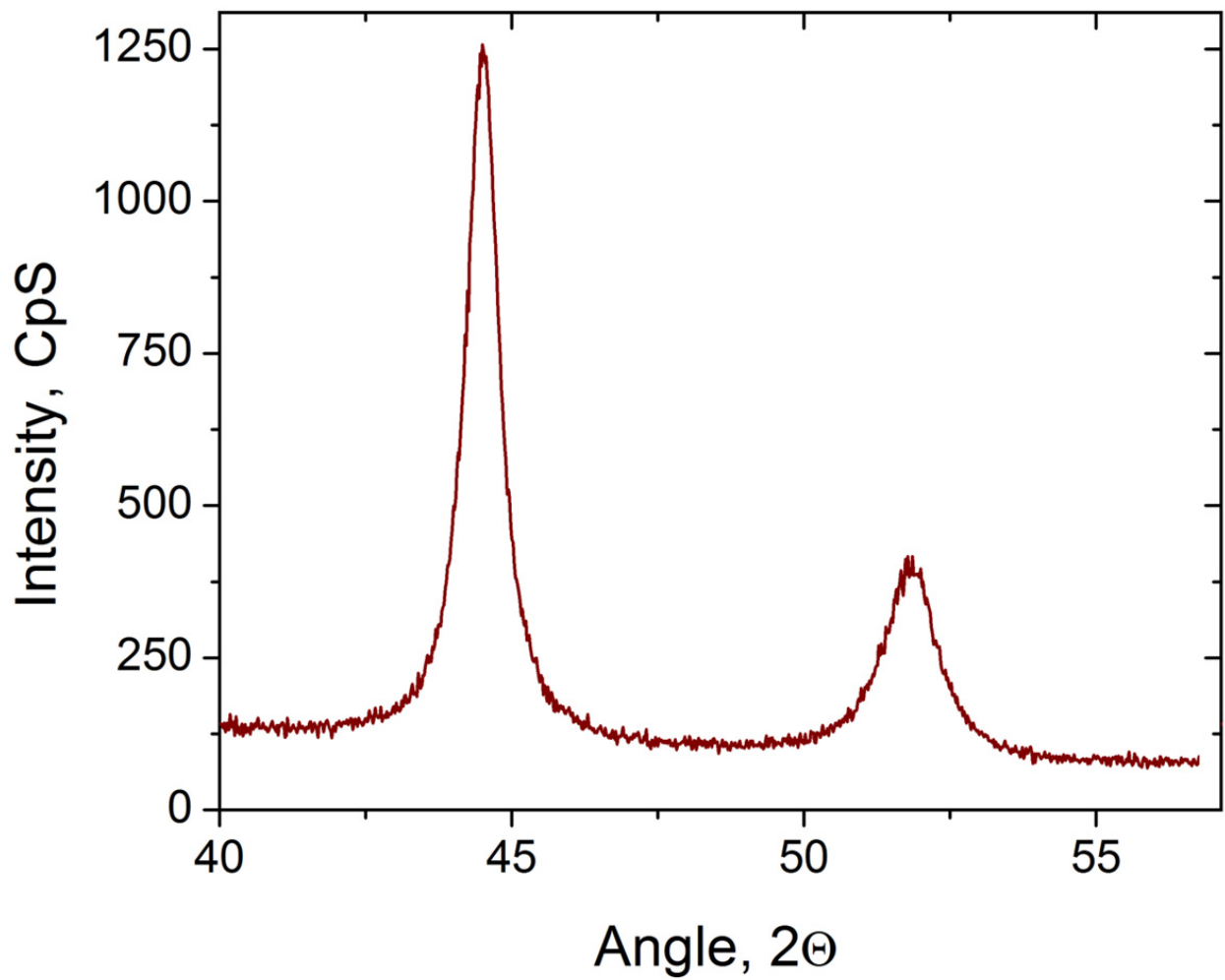

**Figure 11:** Powder XRD pattern of the final nickel particles, obtained after 23.3 h of growth, corresponds to the JCPS 004-0850. The synthesis procedure for growth of uniform Ni nanoparticles fed by matter supplied from the reduction of nickel basic carbonate in polyol, are described in a recent patent [22] and a paper [24].

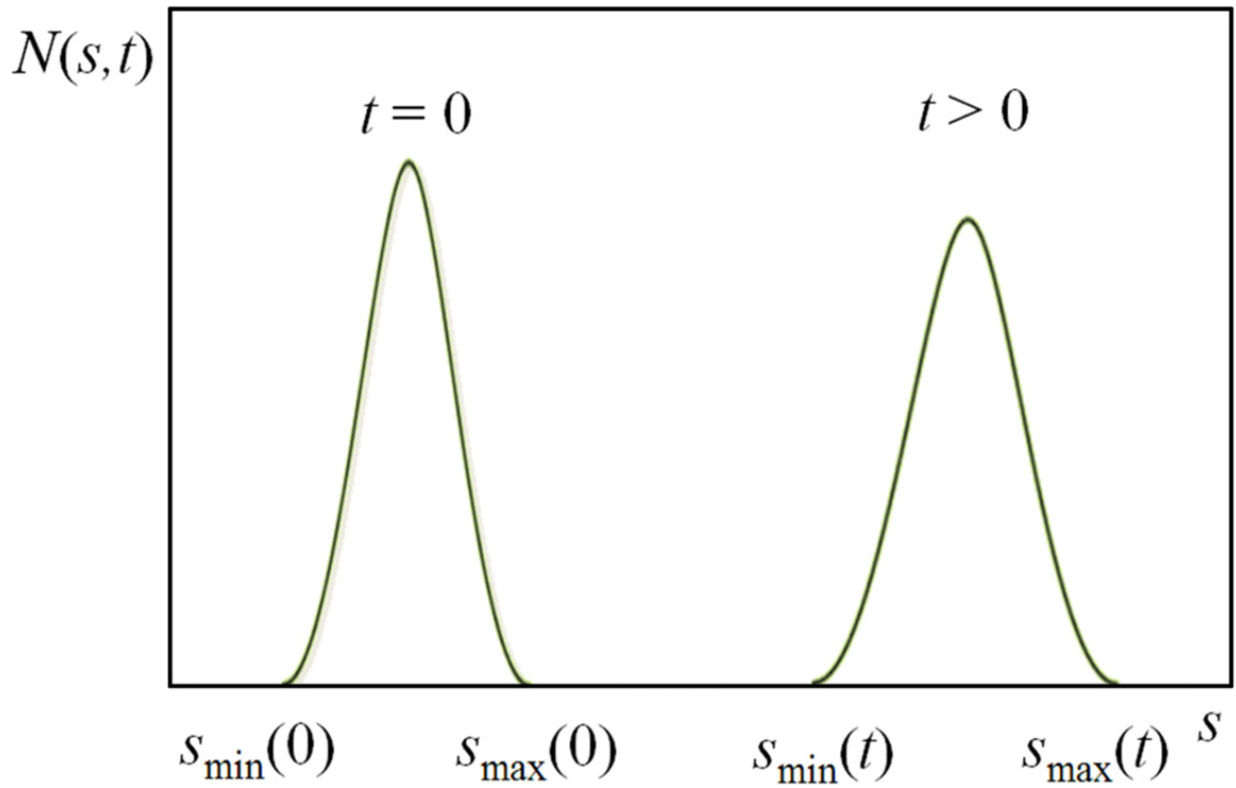

**Figure 12**: Schematic of the growth of the particle size distribution, cf. Eq. (14), starting with the initially seeded particles.

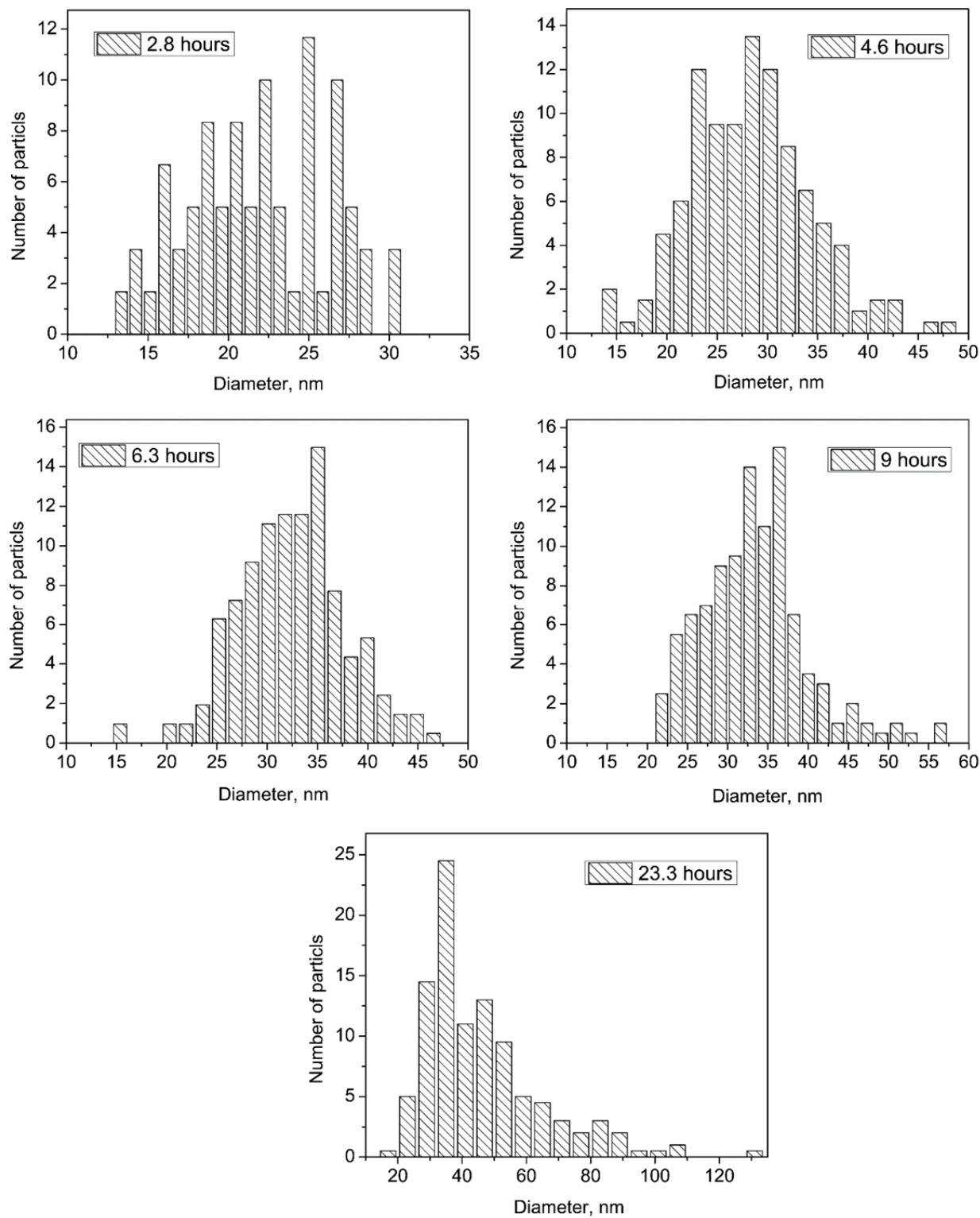

**Figure 13:** Size distribution of nickel particles grown by the diffusional transport over 24 hours in polyol was measured at different times [24].

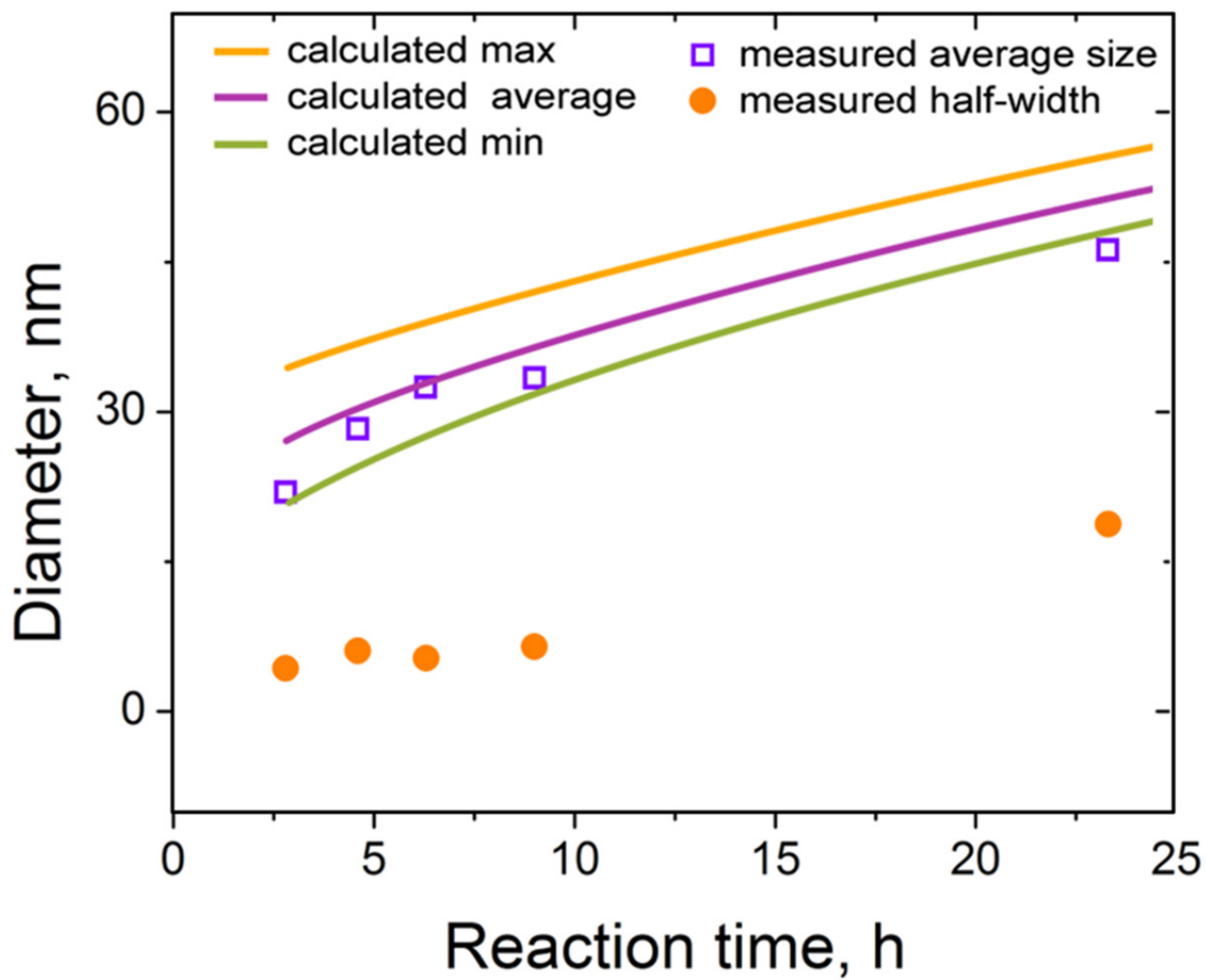

**Figure 14:** Calculated average diameter of the Ni particles, Eq. (16), compared to the experimental values for the average diameter. The max and min quantities defined in Eq. (15) are also drawn. In addition, the half-width values of the measured size distribution are plotted.

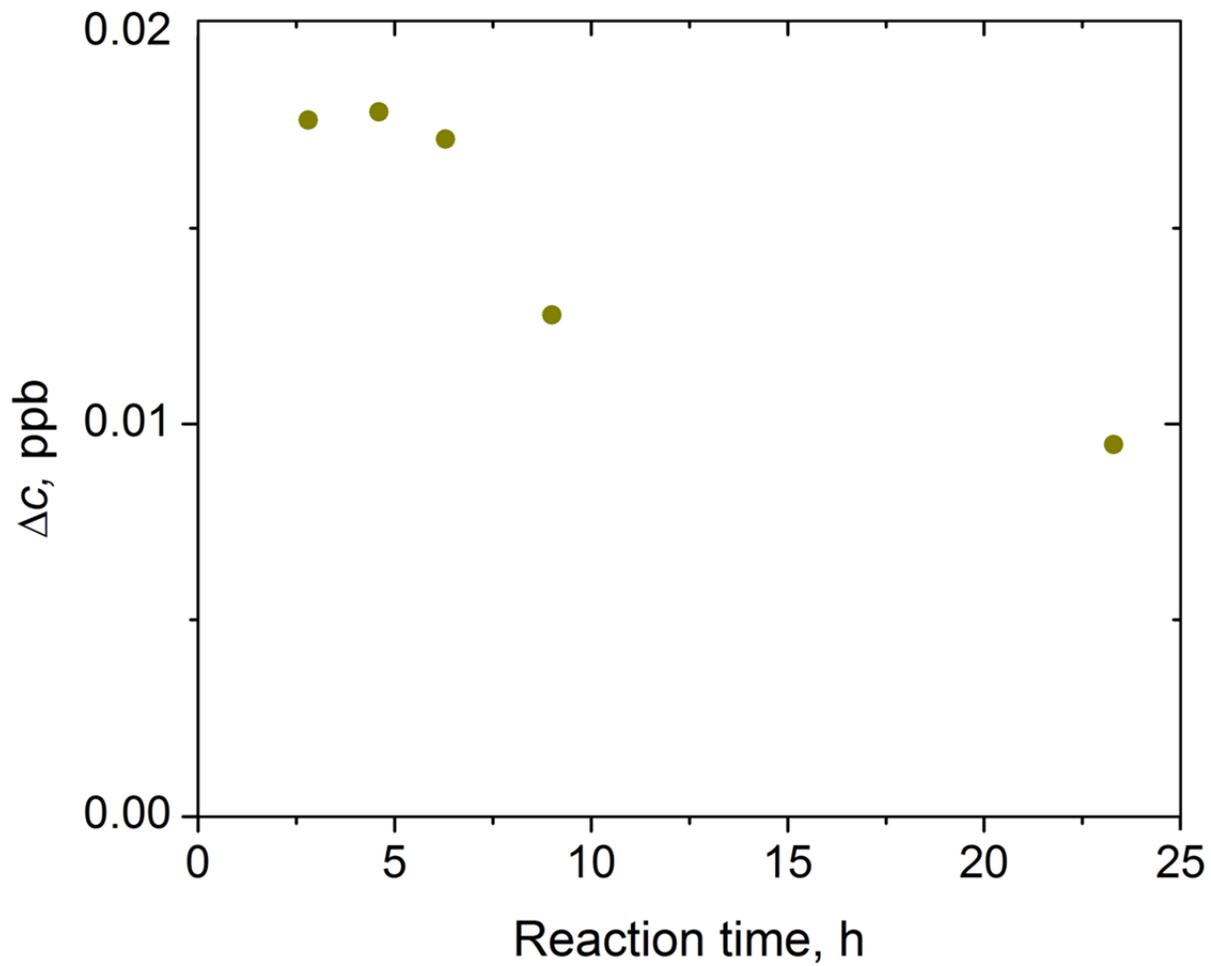

**Figure 15:** Calculated values of the effective excess nickel ion concentration, cf. Eq. (12).

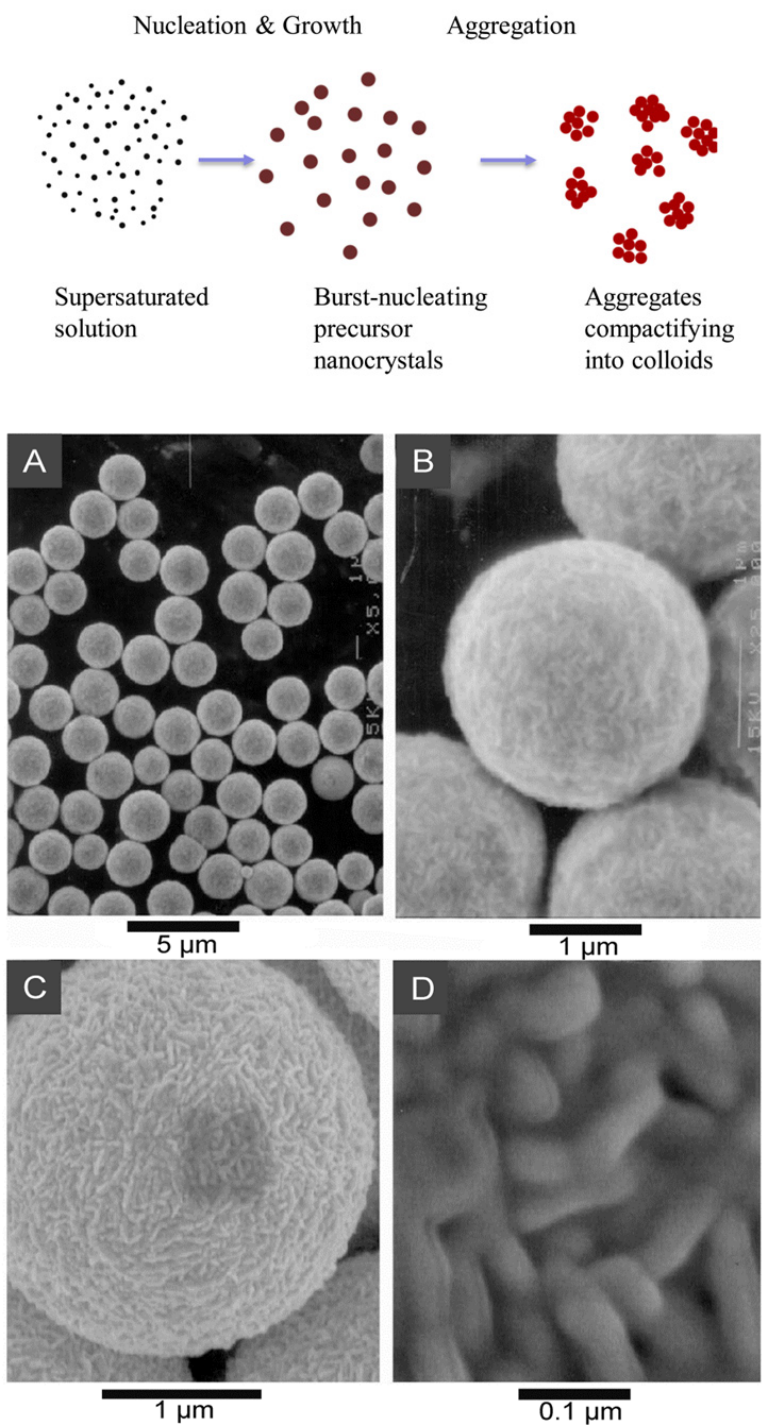

**Figure 16:** Top: Schematic of two-stage synthesis of colloids by aggregation of nanocrystals. Bottom: FESEM mages of gold colloids at increasing magnification in the order of the panel labeling, (A) to (B), (C), and (D).

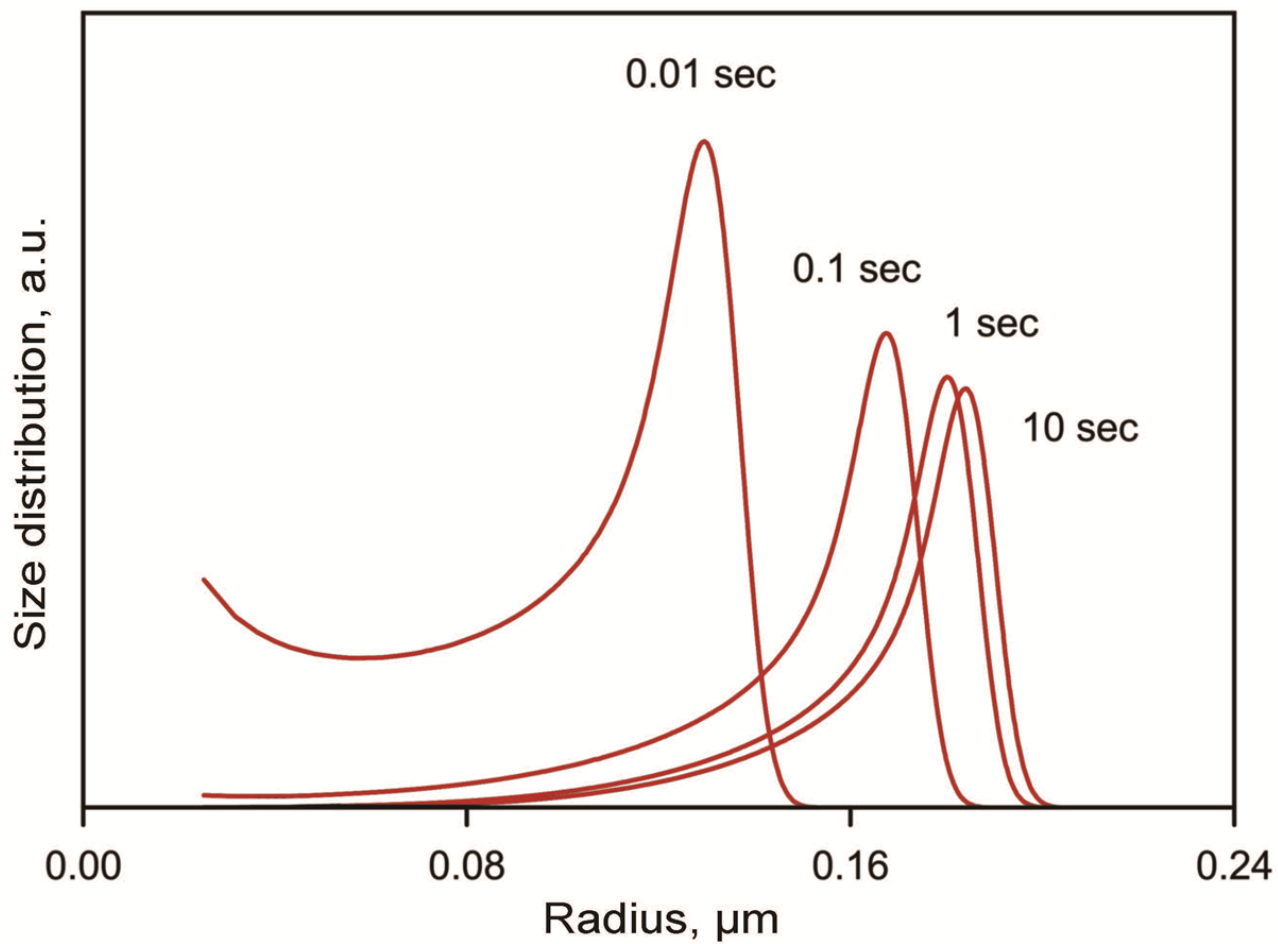

**Figure 17:** Illustration of a growing particle radius distribution for several times, *t*, calculated with model parameters for spherical gold colloids [57, 62].